\documentclass[11pt,leqno]{mf}

\usepackage{amsmath,amsfonts}
\usepackage{amsthm}
\usepackage{fancybox,amssymb}
\usepackage{epsfig}
\usepackage{graphicx}
\usepackage{subfigure}
\usepackage{natbib}

\newtheorem{theorem}{Theorem}[section]

\newtheorem{proposition}{Proposition}[section]
\newtheorem{lemma}{Lemma}[section]
\newtheorem{remark}{Remark}[section]
\newtheorem{definition}{Definition}[section]
\newtheorem{exn}{Example}[section]

\def\e{\mathbf{e}}

\def\u{\mathbf{u}}
\def\P{\mathbf{P}}
\def\0{\mathbf{0}}

\def \K{\mathbf{K}}

\def \E{\mathbb{E}}

\def\={{\!\!\!=\!\!\!}}
\def\proof {{\it Proof:} }
\def\endproof{\hfill $\Box$ \vskip .5cm}

\begin{document}

\title{MEAN-VARIANCE POLICY FOR DISCRETE-TIME CONE CONSTRAINED MARKETS: TIME CONSISTENCY IN EFFICIENCY AND MINIMUM-VARIANCE SIGNED SUPERMARTINGALE MEASURE\thanks{This research work was partially supported by Research Grants Council of Hong Kong under grants 414808, 414610 and 520412, National Natural Science Foundation of China under grant 71201094, and Shanghai Pujiang Program of China under grant 12PJC051. The second author is grateful to the support from the Patrick Huen Wing Ming Chair Professorship of Systems Engineering and Engineering Management.}
}

\author{Xiangyu Cui\thanks{School of Statistics and Management, Shanghai University of Finance and Economics.}, \quad
Duan Li\thanks{Corresponding author. Department of Systems Engineering \& Engineering management, The Chinese University of Hong Kong.} \quad and \quad
Xun Li\thanks{Department of Applied Mathematics, The Hong Kong Polytechnic University.}}

\maketitle

\begin{abstract}
The discrete-time mean-variance portfolio selection formulation, a
representative of general dynamic mean-risk portfolio selection
problems, does not satisfy time consistency in efficiency (TCIE)
in general, i.e., a truncated pre-committed efficient policy may
become inefficient when considering the corresponding truncated
problem, thus stimulating investors' irrational investment
behavior. We investigate analytically effects of portfolio
constraints on time consistency of efficiency for convex cone
constrained markets. More specifically, we derive the
semi-analytical expressions for the pre-committed efficient
mean-variance policy and the minimum-variance signed
supermartingale measure (VSSM) and reveal their close
relationship. Our analysis shows that the pre-committed
discrete-time efficient mean-variance policy satisfies TCIE if and
only if the conditional expectation of VSSM's density (with respect to
the original probability measure) is nonnegative, or once the
conditional expectation becomes negative, it remains at the same
negative value until the terminal time. Our findings indicate that
the property of time consistency in efficiency only depends on the
basic market setting, including portfolio constraints, and this fact
motivates us to establish a general solution framework in
constructing TCIE dynamic portfolio selection problem formulations
by introducing suitable portfolio constraints.
\end{abstract}


\section{Introduction}

In a dynamic decision problem, a decision maker may
face a dilemma when the overall objective for the entire time
horizon under consideration does not conform with a ``local''
objective for a tail part of the time horizon. In the language of
dynamic programming, Bellman's principle of optimality is not
applicable in such situations, as the global and local interests
derived from their respective objectives are not consistent. This
phenomenon has been investigated extensively recently in the
literature of finance and financial engineering under the term  of
\emph{time inconsistency}. In the language of portfolio selection,
when a problem is not time consistent, the (global) optimal
portfolio policy for the entire investment horizon determined at
initial time may not be optimal for a truncated investment problem
at some intermediate time $t$ and for certain realized wealth
level. Investors thus have incentives to deviate from the
global optimal policy and to seek the (local) optimal portfolio
policy, instead, for the truncated time horizon.

As time consistency (or dynamic consistency) is a
basic requirement for dynamic risk measures (see \cite{Rosazza2006}, \cite{BodaFilar2006},
\cite{Artzner2007} and \cite{JobertRogers2008}), all the appropriate dynamic
risk measures should necessarily possess certain functional
structure so that Bellman's principle of optimality is
satisfied. Unfortunately, almost all static risk measures which
investors have been comfortably adopting in practice for decades,
including the variance, VaR (\cite{DuffiePan1997}) and CVaR
(\cite{Uryasev2000}), are not time consistent when being extended
to dynamic situations (\cite{BodaFilar2006}).
Researchers have proposed using the nonlinear expectation
(``g-expectation'') (\cite{Peng1997}) to construct time consistent
dynamic risk measures.

When a dynamic risk measure is time consistent, it not only justifies the mathematical formulation for risk management, but also facilitates the solution process in finding the optimal decision, as the corresponding dynamic mean-risk portfolio selection problem satisfies Bellman's principle of optimality, thus being solvable by dynamic programming (e.g., see \cite{Cherny2010}). When a dynamic risk measure is time inconsistent, the corresponding dynamic mean-risk portfolio selection problem is nonseparable in the sense of dynamic programming, thus generating intractability, or even an insurmountable obstacle in deriving the solution. Consider the dynamic mean-variance portfolio selection problem as an example, as it is the focus of this paper. As the nonseparable structure of the variance term leads to a notoriety of the variance minimization problem, it took almost 50 years to figure out ways to extend the seminal \cite{Markowitz1952}'s static mean-variance formulation to its dynamic counterpart (see \cite{LN} for the discrete-time (multi-period) mean-variance formulation and \cite{ZL} for the continuous-time mean-variance formulation).  The derived dynamic optimal investment policy in \cite{LN} and \cite{ZL} is termed by \cite{BasakChabakauri} as {\it pre-committed} dynamic optimal investment policy, as the (adaptive) optimal
policy is fixed at time 0 to achieve overall optimality for the entire investment horizon.  As the original dynamic mean-variance formulation is not time consistent, the derived {\it pre-committed} dynamic optimal investment policy does not satisfy the principle of optimality and investors have incentive to deviate from such a policy during the investment process in certain circumstances, as revealed in \cite{ZLW1} and \cite{BasakChabakauri}.

There are two major research directions in the literature to alleviate the effects of the time inconsistency of the pre-committed optimal mean-variance policy. To remove the time inconsistency of the pre-committed optimal mean-variance policy, \cite{BasakChabakauri} suggested the so-called time-consistent policy by backward induction in that the investor optimally chooses the (time consistent) policy at any time $t$, on the premise that he has already decided his time consistent policies in the future. \cite{Bjork:2012} extended the formulation in \cite{BasakChabakauri} by introducing state dependent risk aversion and used the backward time-inconsistent control method (see \cite{Bjork:2010}) to derive the corresponding time-consistent policy. \cite{Czichowsky2013} considered the time consistent policies for both discrete-time and continuous-time mean-variance models and revealed the connections between the two. Enforcing a time consistent policy in an inherent time-inconsistent problem undoubtedly incurs a cost, i.e., resulting in a worse mean-variance efficient frontier when compared with the one associated with the pre-committed mean-variance policy, as evidenced from some numerical experiments reported in \cite{WangForsyth2011}. On the other hand, \cite{CuiLiWangZhu2012} relaxed the concept of
time consistency in the literature to ``\emph{time consistency in
efficiency}'' (TCIE)  based on a multi-objective version of the principle of
optimality: The principle of optimality holds if any tail part of an
efficient policy is also efficient for any realizable state at any
intermediate period (\cite{Li1987} and \cite{Li1990}). Note that the essence of the ground breaking work of \cite{Markowitz1952} is to attain an efficiency in portfolio selection by striking a balance between two conflicting objectives of maximizing the expected return and minimizing the investment risk. In this sense, TCIE is nothing but requiring efficiency for any truncated mean-variance portfolio selection problem at every time instant during the investment horizon. \cite{CuiLiWangZhu2012} showed that the dynamic mean-variance problem does not satisfy time consistency in efficiency (TCIE) and developed a TCIE revised mean-variance policy by relaxing the self-financing restriction to allow withdrawal of money out of the market. While the revised policy achieves the same mean-variance pair of the
terminal wealth as the the pre-committed dynamic optimal investment policy does, it also enables investors to
receive a free cash flow stream during the investment
process. The revised policy proposed in \cite{CuiLiWangZhu2012} thus strictly dominates the pre-committed dynamic optimal investment policy.

It is interesting to note that the current literature on time inconsistency has been mainly confined to investigation of time consistent risk measures. While portfolio constraints serve as an important part of the market setting, the literature has been lacking of a study on the effects of portfolio constraints on the property of time consistency and TCIE. Let us consider an extreme situation where only one admissible investment policy is available over the entire investment horizon. In such a situation, no matter whether or not the adopted dynamic risk measure is time consistent, this policy is always optimal and time consistent, as it is the only choice available to investors. Another lesson we could learn is from \cite{WangForsyth2011} where they numerically compared the pre-committed optimal mean-variance policy and the time-consistent mean-variance policy (proposed by \cite{Bjork:2012}) in a continuous-time market with no constraint, with no-bankruptcy constraint or with no-shorting constraint, respectively. They found that with constraints, the efficient frontier generated by the time-consistent mean-variance policy gets closer to the efficient frontier generated by the pre-committed optimal mean-variance policy in the constrained market than in the unconstrained market, i.e., the presence of portfolio constraints may reduce the cost when enforcing a time consistent policy in an inherent time-inconsistent problem. Based on the above recognition, it is our purpose to study in this paper analytically the impact of convex cone-type portfolio constraints on TCIE in a discrete-time market. Our analysis reveals an ``if and only if'' relationship between TCIE and the conditional expectation of the density of the minimum-variance signed supermartingale measure (with respect to
the original probability measure). As our finding indicates
that the property of time consistency in efficiency only depends on the basic market setting,
including portfolio constraints, we further establish a general solution framework in
constructing TCIE dynamic portfolio selection problem formulations by introducing suitable
portfolio constraints.

The main theme and the contribution of this paper is to address and answer the following question: Given a financial market with its return statistics known, what are the cone constraints on portfolio policies or what additional cone constraints are needed to be introduced such that the derived optimal portfolio policy is TCIE.  The paper is thus organized to present this story line with the following key points in achieving this overall research goal. For a general class of discrete-time convex cone constrained markets, we derive analytically the pre-committed discrete-time efficient mean-variance policy using duality theory and dynamic programming (Section 2). Theorem \ref{prop:cone-constained} fully characterizes the distinct features of this policy and, in particular, reveals that the optimal policy is a two-piece linear function of the current wealth, while the time-varying breaking point of the two pieces is determined by a deterministic threshold wealth level. We then discuss the necessary and sufficient conditions for the pre-committed efficient policy to be TCIE (Section 3). Theorem \ref{prop:truncated-efficency} specifies the behavior pattern of TCIE policies for both cases below and above the threshold wealth level. We define and derive the minimum-variance signed supermartingale measure (VSSM) for cone constrained markets and reveal its close relationship with TCIE (Section 4). More specifically, we show in Theorem \ref{thm:TCIE-VSSM} that the pre-committed efficient mean-variance policy satisfies TCIE if and only if the conditional expectation of VSSM's density (respect to the original probability measure) is nonnegative, or once the conditional expectation becomes negative, it remains at the same negative value until the terminal time. We finally answer the question how to completely eliminate time inconsistency in efficiency by introducing additional cone constraints to the market (Section 5). Theorem \ref{prop:condition-TCIE} can be viewed as the culmination of all the results in this paper, in which a constructive framework in achieving TCIE is established through identifying a convex cone for constraining portfolios such that its dual cone includes the given expected excess return vector of the market under consideration. In order to make our presentation clear, we have placed all the proofs in the appendix.

\section{Optimal mean-variance policy in a discrete-time cone constrained market}

The capital market of $T$ time periods under consideration consists of $n$ risky assets with random rates of returns and one riskless asset with a deterministic rate of return. An investor with an
initial wealth $x_0$ joins the market at time $0$ and allocates
his wealth among these $(n+1)$ assets. He can reallocate his wealth
among the $(n+1)$ assets at the beginning of each of the following
$(T-1) $ consecutive time periods. The deterministic rate of
return of the riskless asset at time period $t$ is denoted by
$s_t>0$ and the rates of return of the risky assets at time period
$t$ are denoted by a vector $\mathbf{e}_t=[e_t^1, \cdots,
e_t^n]'$, where $e_t^i$ is the random return of asset $i$ at time
period $t$ and the notation $'$ denotes the transpose operation. It is assumed in this paper that vectors
$\mathbf{e}_t$, $t = 0, 1, \cdots, T - 1$, are statistically
independent with mean vector
$\mathbb{E}[\mathbf{e}_t] = [\mathbb{E}[e_t^1], \cdots,
\mathbb{E}[e_t^n]]'$ and positive definite covariance matrix,
\[
\mbox{Cov}\left( \mathbf{e}_t\right) =\left[
\begin{array}{ccc}
\sigma _{t,11}  & \cdots & \sigma _{t,1n}\\
\vdots & \ddots & \vdots \\
\sigma _{t,1n} & \cdots & \sigma _{t,nn}
\end{array}
\right]\succ 0.
\]
Assume that all the random vectors, $\e_t$, $t=0,1,\cdots,T-1$, are defined in a filtrated probability space $(\Omega,\mathcal{F}_T, \{\mathcal{F}_t\}, P)$, where $\mathcal{F}_t=\sigma\left(\e_0,\e_1,\cdots,\e_{t-1}\right)$ and $\mathcal{F}_0$ is the trivial $\sigma$-algebra over $\Omega$. Therefore, $\E[\cdot|\mathcal{F}_0]$ is just the unconditional expectation $\E[\cdot]$. Let $x_t$ be the wealth of the investor at the beginning of the
$t$-th time period, and $u_t^i$, $i = 1, 2, \cdots, n$, be the dollar
amount invested in the $i$th risky asset at the beginning of the
$t$-th time period. The dollar amount invested in the riskless asset at the
beginning of the $t$-th time period is then equal to $x_t -
\sum_{i=1}^n u^i_t$. It is assumed that the admissible investment strategy $\mathbf{u}_t=[u_t^1,u_t^2, \cdots, u_t^n]'$ is an $\mathcal{F}_t$-measurable Markov control, i.e., $\u_t\in\mathcal{F}_t$, and the realization of $\u_t$ is restricted to a deterministic and non-random convex cone $\mathcal{A}_t\subseteq\mathbb{R}^n$. Such cone type constraints are of wide application in practice to model regulatory restrictions, for example, restriction of no short selling and restriction for non-tradeable assets. Cone type constraints are also useful to represent portfolio restrictions, for example, the holding of the first asset must be no less than the second asset, which can be generally expressed by $\mathcal{A}_t=\{\u_t\in\mathbb{R}^n|A \u_t\geq 0,~A\in\mathbb{R}^{m\times n}\}$ (see \cite{Cuoco1997} and \cite{Napp2003} for more details).

An investor of
mean-variance type seeks the best admissible investment strategy,
$\{\mathbf{u}_t^*\}\mid_{t=0}^{T-1}$, such that the variance of the
terminal wealth,  $\mbox{Var}(x_T)$, is minimized subject to that
the expected terminal wealth, $\mathbb{E}[x_T]$, is fixed at a
preselected level $d$,
\begin{equation}\label{prob:cone_constrained}
\begin{array}{rl}
(P(d)): & \quad \left\{\begin{array}{rl}
\min & \mbox{Var}(x_T)\equiv \mathbb{E}\big[(x_T-d)^2 \big], \\ [2mm]
\mbox{\rm s.t.} & \mathbb{E}[x_T] = d, \\ [2mm]
& x_{t+1} =s_tx_t+\mathbf{P}_t'\mathbf{u}_t, \\ [2mm]
& \mathbf{u}_t\in\mathcal{A}_t, \quad t = 0, 1, \cdots, T-1,
\end{array}\right.
\end{array} \nonumber
\end{equation}
where
\[
\P_t=\left[ P_t^1,P_t^2, \ldots, P_t^n\right] ^{\prime }=\left[ (e_t^1-s_t),(e_t^2-s_t), \ldots, (e_t^n-s_t)\right]'
\]
is the vector of the excess rates of returns. It is easy to see that $\P_t$ and $\u_t$ are independent, $\{x_t\}$ is an adapted Markovian process and $\mathcal{F}_t = \sigma(x_t)$.

\begin{remark} Varying parameter $d$ in $(P(d))$ from $-\infty$ to $+\infty$ yields the minimum variance set in the mean-variance space. Furthermore, as setting $d$ equal to $\prod_{i=0}^{T-1}s_ix_0$ in $(P(d))$ gives rise to the minimum variance point, the upper branch of the minimum variance set
corresponding to the range of $d$ from
$\prod_{i=0}^{T-1}s_ix_0$ to $+\infty$ characterizes the efficient frontier in
the mean-variance space which enables investors to recognize the
trade-off between the expected return and the risk, thus helping
them specify their preferred expected terminal wealth.
\end{remark}

Note that condition $\mbox{Cov}\left( \mathbf{e_t}\right) \succ 0$ implies the positive definiteness of the second moment of
$(s_t,\e_t')'$. The following is then
true for $t=0,1,\cdots,T-1$:
\[
\begin{array}{rl}
& \left[\begin{array}{cc} s_t^2 & s_t\E[\P_t'] \\
s_t\E[\P_t] & \E[\P_t\P_t']
 \end{array}\right] \\ [5mm]
\= & \left[\begin{array}{cccc} 1 & 0 & \cdots & 0 \\
 -1 & 1 &\cdots &0 \\
 \cdots & \cdots & \cdots &\cdots \\
 -1 & 0 &\cdots &1
 \end{array}\right]
 \left[\begin{array}{cc} s_t^2 & s_t\E[\e_t'] \\
s_t\E[\e_t] & \E[\e_t\e_t']
 \end{array}\right]\left[\begin{array}{cccc} 1 & -1 &\dots &-1 \\
 0 & 1  & \cdots & 0 \\
 \cdots & \cdots & \cdots & \cdots \\
 0 & 0  & \cdots & 1
\end{array}\right]\succ 0,
\end{array}
\]
which further implies
\[
\begin{array}{rl}
\E[\P_t\P_t']\succ 0,& \forall \ t=0,1,\cdots,T-1, \\ [2mm]
s_t^2(1-\E[\P_t']\E^{-1}[\P_t\P_t']\E[\P_t])> 0, & \forall \ t=0,1,\cdots,T-1.
\end{array}
\]

Constrained dynamic mean-variance portfolio selection problems
with various constraints have been attracting increasing attention
in the last decade, e.g., \cite{LZL}, \cite{ZLW2},
\cite{Bielecki2005},  \cite{SunWang}, \cite{LabbeHeunis} and
\cite{CzichowskySchweizer2010}. Recently,
\cite{CzichowskySchweizer2013} further considered cone-constrained
continuous-time mean-variance portfolio selection with price
processes being semimartingales.

\begin{remark} In this section, we will use
duality theory and dynamic programming to derive the discrete-time
efficient mean-variance policy analytically in convex cone
constrained markets. We will demonstrate that the optimal
mean-variance policy is a two-piece linear function of the current
wealth level, which represents an extension of the result in
\cite{CuiGaoLiLi2012} for discrete-time markets under the
no-shorting constraint (a special convex cone) and a discrete-time
counterpart of the policy in \cite{CzichowskySchweizer2013}.
\end{remark}

We define the following two deterministic functions, $h_t^{+}(\K_t)$ and $h_t^{-}(\K_t)$, on $\mathbb{R}^n$ for $t=0,1,\dots,T-1$,
\begin{equation}
h_t^{\pm}(\K_t) = \E\left[C_{t+1}^+\Big( 1\mp\P_t^{\prime}\K_t\Big)^21_{\{\P_t^{\prime}\K_t\leq \pm1\}}+C_{t+1}^-\Big( 1\mp\P_t^{\prime}\K_t\Big)^21_{\{\P_t^{\prime}\K_t> \pm1\}}\right],\label{eq:h-function}\\
\end{equation}
with terminal condition $C_T^{+}=C_T^{-}=1$, and denote their deterministic minimizers and optimal values, respectively, as
\begin{align}
&\K_t^{\pm} = \displaystyle\arg\!\min_{\K_t\in\mathcal{A}_t}\E\left[C_{t+1}^+\Big( 1\mp\P_t^{\prime}\K_t\Big)^21_{\{\P_t^{\prime}\K_t\leq \pm1\}}+C_{t+1}^-\Big( 1\mp\P_t^{\prime}\K_t\Big)^21_{\{\P_t^{\prime}\K_t> \pm1\}}\right],\label{eq:opt-K}\\
&C_t^{\pm} =\E\left[C_{t+1}^+\Big( 1\mp\P_t^{\prime}\K_t^{\pm}\Big)^21_{\{\P_t^{\prime}\K_t^{\pm}\leq \pm1\}}+C_{t+1}^-\Big( 1\mp\P_t^{\prime}\K_t^{\pm}\Big)^21_{\{\P_t^{\prime}\K_t^{\pm}> \pm1\}}\right]. \label{eq:C-1}
\end{align}
As will be seen later in the paper, functions $\K_t^{\pm}$ and $C_t^{\pm}$ appear in the optimal policy for problem $(P(d))$. The following lemma is important in deriving our main result in this paper.
\begin{lemma}\label{lem:K-C} \sl
For $t=0,1,\dots,T-1$, the following properties hold,
\begin{align}
&C_t^{\pm}=\E\left[C_{t+1}^+\Big( 1\mp\P_t^{\prime}\K_t^{\pm}\Big)1_{\{\P_t^{\prime}\K_t^{\pm}\leq \pm1\}}+C_{t+1}^-\Big( 1\mp\P_t^{\prime}\K_t^{\pm}\Big)1_{\{\P_t^{\prime}\K_t^{\pm}> \pm1\}}\right],\label{eq:C}\\
\nonumber &0<C_t^{\pm}\leq C_{t+1}^{\pm}.
\end{align}
Furthermore, $C_t^{\pm}=C_{t+1}^{\pm}$ if and only if $\K_t^{\pm}=\0$ (Notation $\0$ denotes the $n$-dimensional zero vector).
\end{lemma}

Note that Lemma \ref{lem:K-C} reduces the piecewise quadratic form of $C_t^{\pm}$ in (\ref{eq:C-1}) to a piecewise linear one in (\ref{eq:C}). We can adopt Lagrangian duality and dynamic programming to solve
problem $(P(d))$.
\begin{theorem}\label{prop:cone-constained} \sl
Define $\rho_t =\prod_{\ell=t}^{T-1}s_\ell$ {\rm (with~}$\prod_{i\in \emptyset}f_i$ being set to 1{\rm )}. When both $d>\rho_0x_0$ and $C_0^+=1$ hold, or both $d<\rho_0x_0$ and $C_0^-=1$ hold, problem $(P(d))$ does not have a feasible solution.

Under the assumption that problem $(P(d))$ is feasible, its optimal investment policy can be expressed by the following deterministic piecewise linear function of wealth level $x_t$,
\begin{align}\label{eq:opt-u-star}
\u_t^\star(x_t) =s_t\K_t^+\big((d-\mu^{\star})\rho_t^{-1}-x_t\big)1_{\{d-\mu^{\star}\geq \rho_t x_{t}\}}-s_t\K_t^-\big((d-\mu^{\star})\rho_t^{-1}-x_t\big)1_{\{d-\mu^{\star}< \rho_t x_{t}\}},\\
\nonumber t=0,1,\dots,T-1,
\end{align}
where
\begin{align}\label{eq:opt-mu}
\mu^{\star} &= \displaystyle\frac{d-\rho_0 x_{0}}{1-(C_0^+)^{-1}}1_{\{d\geq \rho_0 x_{0}\}}+\frac{d-\rho_0 x_{0}}{1-(C_0^-)^{-1}}1_{\{d< \rho_0 x_{0}\}}.
\end{align}
Moreover, the minimum variance set is given as
\[
\mbox{\rm Var}(x_T)=\frac{C_0^+\big(\mathbb{E}[x_T]-\rho_0
x_{0}\big)^2}{1-C_0^+}1_{\{\E[x_T]\geq
\rho_0x_0\}}+\frac{C_0^-\big(\mathbb{E}[x_T]-\rho_0
x_{0}\big)^2}{1-C_0^-}1_{\{\E[x_T]<
\rho_0x_0\}},
\]
and the mean-variance efficient frontier, which is the upper branch of the minimum variance set, is expressed as
\begin{equation}\label{eq:eff-frontier}
\mbox{\rm Var}(x_T)=\frac{C_0^+\big(\mathbb{E}[x_T]-\rho_0
x_{0}\big)^2}{1-C_0^+},\quad \mbox{for}\quad \E[x_T]\geq
\rho_0x_0.
\end{equation}
\end{theorem}
Note that every point on the lower branch of the minimum variance set corresponding to $d$ $<$ $\rho_0x_0$ is dominated by the minimum variance point with $\E[x_T]$ =
$\rho_0x_0$ and $\mbox{\rm Var}(x_T)$ = 0. Although the cases with $d$ $<$ $\rho_0x_0$  do not make sense from an economic point of view for the entire investment horizon, we do need this explicit expression for the lower branch of the minimum variance set for our later discussion in the paper. As we demonstrate later in the paper, the pre-committed investment policy is not time consistent in efficiency. Thus, applying the pre-committed mean-variance policy for a truncated time horizon could result in an inefficient mean-variance pair which falls onto the lower branch of the minimum variance set for the truncated time horizon. Time inconsistency in efficiency hides behind this kind of phenomena which is not economically sensible. The purpose of this paper is to devise a solution scheme to eliminate time inconsistency in efficiency, thus removing this kind of phenomena with no economic sense.

\begin{remark}
  Theorem \ref{prop:cone-constained} reveals that the optimal investment policy is a two-piece linear function with respect to the investor's current wealth level and this finding represents the discrete-time counterpart of the result in \cite{CzichowskySchweizer2013} for continuous-time. In Section 5, we will also demonstrate that the result in Theorem \ref{prop:cone-constained} is also an extension of the result in \cite{CuiGaoLiLi2012} for multiperiod mean-variance formulation with no-shorting constraint.
  \end{remark}

When $d\geq \rho_0 x_{0}$ and $C_0^+<1$ hold, the optimal investment policy $\u_t^\star$, $t=0,1,\cdots, T-1$, in (\ref{eq:opt-u-star}) is efficient, which we term as a pre-committed efficient mean-variance policy following \cite{BasakChabakauri}. When $d=\rho_0x_0$, the optimal investment policy is achieved by $\u_t^\star=\0$, i.e., investor invests all his wealth in the riskless asset, which is exactly the minimum variance policy. When $d< \rho_0 x_{0}$ and $C_0^-<1$ hold, the optimal investment policy of $(P(d))$, $\u_t^\star$, $t=0,1,\cdots, T-1$, in (\ref{eq:opt-u-star}) is inefficient.

\begin{remark}
By setting $\mathcal{A}_t=\mathbb{R}^n$, $t=0,1,\dots,T-1$, the pre-committed discrete-time efficient mean-variance policy in (\ref{eq:opt-u-star}) reduces to the one in the unconstrained market (\cite{LN}),
\begin{align*}
\u_t^\star(x_t)=s_t\big((d-\mu^{\star})\rho_t^{-1}-x_t\big)\E^{-1}\left[\P_t\P_t^{\prime}\right]\E\left[\P_t\right], ~~t=0,1,\dots,T-1,
\end{align*}
where
\begin{align*}
\mu^{\star} &= \displaystyle\frac{d-\rho_0 x_{0}}{1-\prod_{i=0}^{T-1}(1-\E\left[\P_i^\prime\right]\E^{-1}\left[\P_i\P_i^{\prime}\right]\E\left[\P_i\right])^{-1}}.
\end{align*}

The major differences between the pre-committed efficient mean-variance policies in a cone constrained market and in the unconstrained market lie in the following three aspects. First, in a cone constrained market, problem $(P(d))$ may become infeasible, while feasibility is never an issue for the mean-variance portfolio selection in unconstrained markets. Second, when $\P_t$, $t=0,1,\cdots,T-1$, are identically distributed, $\E^{-1}[\P_t\P_t^\prime]\E[\P_t]$, $t=0,1,\cdots,T-1$, take the same value, which implies that investors hold a unique risky portfolio for any time period in an unconstrained market, which is also independent of the wealth level of the investor. In a cone constrained market, however, the investor may hold two different risky portfolios, $\K_t^+$ and $\K_t^-$, while $\K_t^+$ and $\K_t^-$ are in general different for the same time $t$. A key observation thus is that the investor may switch his risky position according to his current wealth level. Third, in a cone constrained market, although the excess rates of return of risky assets, $\P_t$, $t = 0, 1, \cdots, T - 1$, are statistically
independent, the future $\P_\tau$, $\tau>t$, may influence the current risky portfolios, $\K_t^+$ and $\K_t^-$, through parameters $C_t^+$ and $C_t^-$, which implies that the independent structure of the optimal risky portfolio holding (rooted from the independent assumption of the random rate of return) is destroyed by the presence of constraints. Thus, in general, $\K_t^\pm\neq \K_s^\pm$ when $t\neq s$, which implies further that the risky positions of the investor are not time-invariant anymore. In summary, we can conclude that, in a cone-constrained market, the risky positions are both state-dependent and time-dependent.
\end{remark}

\section{Conditions for Time Consistency in Efficiency of the Pre-committed Efficient Mean-Variance Policy}

We check now the performance of the pre-committed
optimal mean-variance policy
$\{\mathbf{u}_t^\star\}\mid_{t=0}^{T-1}$ derived for the entire investment time horizon given in
(\ref{eq:opt-u-star}) of the last section in truncated time periods. More specifically, we would like to examine the efficiency of $\{\mathbf{u}_t^\star\}\mid_{t=0}^{T-1}$ in shorter time periods and develop conditions under which $\{\mathbf{u}_t^\star\}\mid_{t=0}^{T-1}$ remains efficient all the time. Let us consider the following truncated mean-variance problem for any realized wealth $x_k$ in time period $k$,
\begin{equation}\label{prob:cone_constrained_truncated}
\begin{array}{rl}
(P_k(d_k) \mid x_k): & \quad \left\{\begin{array}{rl}
\min & \mbox{Var}(x_T)= \mathbb{E}\big[(x_T-d_k)^2 \big] \\ [2mm]
\mbox{\rm s.t.} & \mathbb{E}[x_T] = d_k, \\ [2mm]
& x_{t+1} =s_tx_t+\mathbf{P}_t'\mathbf{u}_t, \\ [2mm]
& \mathbf{u}_t\in\mathcal{A}_t, \quad t = k, k+1, \cdots, T-1,\\[2mm]
& x_k \mbox{ is known,}
\end{array}\right.
\end{array} \nonumber
\end{equation}
where $d_k$ is a preselected level of the expected final wealth for the truncated mean-variance problem. As problem $(P_k(d_k) \mid x_k)$ is of the same structure of problem $(P(d))$, based on Theorem \ref{prop:cone-constained}, the corresponding optimal policy of $(P_k(d_k))$ is given by
\begin{align}
\nonumber\tilde{\u}_t^\star(x_t \mid d_k)=s_t\K_t^+\big((d_k-\mu^{\star}_k)\rho_t^{-1}-x_t\big)1_{\{d_k-\mu^{\star}_k\geq \rho_t x_{t}\}}-s_t\K_t^-\big((d_k-\mu^{\star}_k)\rho_t^{-1}-x_t\big)1_{\{d_k-\mu^{\star}_k< \rho_t x_{t}\}},\\
t=k,k+1,\dots,T-1,\label{eq:opt-u-star-truncated}
\end{align}
where
\begin{align}\label{eq:opt-mu-truncated}
\mu^{\star}_k &= \displaystyle\frac{d_k-\rho_k x_{k}}{1-(C_k^+)^{-1}}1_{\{d_k\geq \rho_k x_{k}\}}+\frac{d_k-\rho_k x_{k}}{1-(C_k^-)^{-1}}1_{\{d_k< \rho_k x_{k}\}}.
\end{align}
Evidenced from our discussion on $(P(d))$, the solution to $(P_k(d_k) \mid x_k)$ is mean-variance efficient at $x_k$ if and only if $d_k$ $\geq$ $\rho_k x_k$.

Now we consider the following inverse optimization problem of $(P_k(d_k) \mid x_k)$: For any $x_k$,
$k=1,2,\dots, T-1$, find an expected final wealth level $d_k$  such that the truncated pre-committed
optimal mean-variance policy
$\mathbf{u}_t^\star(x_t)~(t=k,k+1,\dots,T-1)$, with $x_t$ = $x_k$, specified in
(\ref{eq:opt-u-star}) solves $(P_k(d_k) \mid x_k)$. We call such a $d_k$ an {\it induced expected final wealth level} by the pre-committed policy at $x_k$. It becomes evident now that if for some $x_k$, $k=1,2,\dots, T-1$, the induced $d_k$ is less than $\rho_k x_k$, then the truncated pre-committed
optimal mean-variance policy
$\mathbf{u}_t^\star(x_t)~(t=k,k+1,\dots,T-1)$, with $x_t$ = $x_k$, is inefficient for the truncated mean-variance problem from stage $k$ to $T$ with given $x_k$.

\begin{definition}
An efficient solution of $(P(d))$, $\{\u_t^\star(x_t)\} \mid_{t=0}^{T-1}$, is time consistent in efficiency (TCIE) if for all wealth $x_k$ in time period $k$, $k$ = 1, $\ldots$, $T - 1$, the induced expected final wealth level $d_k$ always satisfies $d_k$ $\geq$ $\rho_k x_k$, such that $\{\u_t^\star(x_t)\} \mid_{t=k}^{T-1}$ solves $(P_k(d_k) \mid x_k)$.
\end{definition}
In plain language, a globally mean-variance efficient solution is TCIE if it is also locally mean-variance efficient for every intermediate stage and every possible realizable state (wealth level $x_t$).

\begin{remark} Note that the above definition of TCIE shares the same spirit as the one in \cite{CuiLiWangZhu2012}. However, the current one is defined in terms of the induced expected final wealth, while the one in \cite{CuiLiWangZhu2012} is defined in terms of the induced trade off between the variance and the expectation of the terminal wealth.
 \end{remark}

 \begin{remark}
 Note also that insisting time consistency of $\{\u_t^\star(x_t)\} \mid_{t=0}^{T-1}$ implies that $\{\u_t^\star(x_t)\} \mid_{t=k}^{T-1}$ solves $(P_k(d) \mid x_k)$ for any realized wealth $x_k$ in every time period $k$, $k$ = 1, $\ldots$, $T - 1$.
 \end{remark}

\begin{remark}
\cite{CuiLiWangZhu2012} showed that discrete-time mean-variance portfolio selection problem is not time consistent in efficiency (TCIE) in unconstrained markets. When the investor's wealth level exceeds a deterministic level determined by the market setting, he may become irrational to minimize both the mean and the variance when continuing applying the pre-committed efficient policy. We will check in this section whether discrete-time mean-variance portfolio selection problem in cone constrained markets is also not TCIE.
\end{remark}

 Note that the truncated minimum variance policy is always the minimum variance policy of the corresponding truncated mean-variance problem. Therefore, we only need to check whether the truncated pre-committed efficient policy (expect for the minimum variance policy), $\u_t^\star(x_t)$, $t=k,k+1,\dots,T-1$, is efficient or not with respect to the corresponding truncated mean-variance problem.

\begin{theorem}\label{prop:truncated-efficency} \sl
The truncated pre-committed efficient mean-variance policy (except for the minimum variance policy), $\u_t^{\star}(x_t), t=k,k+1,\dots,T-1$, is also an efficient policy of the truncated problem $(P(d_k) \mid x_k)$, if and only if
\begin{align*}
\mbox{\rm (i)  } d-\mu^{\star}\geq \rho_kx_k, \quad\mbox{ or }\quad \mbox{\rm (ii)  } d-\mu^{\star}< \rho_kx_k,~~ C_k^-=1.
\end{align*}
\end{theorem}

Condition (i) in Theorem \ref{prop:truncated-efficency} for the efficiency of the truncated pre-committed efficient mean-variance policy at time $k$ can be interpreted as a threshold condition for $x_k$,
\begin{align*}
x_k \leq \rho_k^{-1}(d-\mu^\star)=\rho_k^{-1}\frac{d-C_0^+\rho_0x_0}{1-C_0^+},
\end{align*}
which is similar to the result of Proposition 3.1 in \cite{CuiLiWangZhu2012}. On the other hand, note from the last statement in Lemma \ref{lem:K-C}, if $C_k^-$ becomes 1, then all $C_t^-$ with $k < t \leq T-1$ will remain 1, implying $\K^-_t$ = 0, $k \leq t \leq T-1$.
Therefore, condition (ii) in Theorem \ref{prop:truncated-efficency} can be interpreted as follows: Once the wealth level at time $k$ exceeds the deterministic level, $\rho_k^{-1}(d-\mu^\star)$, investor switches to adopt the minimum variance policy (to invest all his wealth in the riskless asset). With the help of Eq. (\ref{eq:opt-u-star}), under both conditions the investor either holds portfolio $\K_k^+$ or only invests in riskless asset. Thus, we term $\K_k^+$ as \emph{efficient risky portfolio}. In contrast, when $d-\mu^{\star}< \rho_kx_k$, $C_k^-<1$, the truncated pre-committed efficient mean-variance policy is inefficient and the corresponding portfolio $\K_k^-$ is thus termed as \emph{inefficient risky portfolio}.

Based on Proposition 3.1 and the definition of time consistency in efficiency, the following lemma for TCIE of the pre-committed efficient mean-variance policy is apparent.

\begin{lemma}\label{lem:TCIE-condition}
The pre-committed efficient mean-variance policy (except for the minimum variance policy) is TCIE if and only if condition (i) or condition (ii) holds for all possible $x_t$ achieved by pre-committed efficient mean-variance policy and for all $t=1,2,\cdots,T-1$.
\end{lemma}

\begin{remark} The following proposition betters our understanding further for investigating the possibility in achieving TCIE.

\begin{proposition}\label{prop:probability-issue} \sl
Adopting the pre-committed efficient mean-variance policy at time $t$ yields the following conditional probabilities,
\begin{align*}
&Pr\left((d-\mu^\star)\geq\rho_{t+1}x_{t+1}\big|(d-\mu^\star)>\rho_{t}x_{t}\right)=Pr\left(\P_t^\prime\K_t^+\leq 1\right),\\
&Pr\left((d-\mu^\star)<\rho_{t+1}x_{t+1}\big|(d-\mu^\star)>\rho_{t}x_{t}\right)=Pr\left(\P_t^\prime\K_t^+>1\right),\\
&Pr\left((d-\mu^\star)\geq\rho_{t+1}x_{t+1}\big|(d-\mu^\star)<\rho_{t}x_{t}\right)=Pr\left(\P_t^\prime\K_t^-\leq -1\right),\\
&Pr\left((d-\mu^\star)<\rho_{t+1}x_{t+1}\big|(d-\mu^\star)<\rho_{t}x_{t}\right)=Pr\left(\P_t^\prime\K_t^-> -1\right),\\
&Pr\left((d-\mu^\star)=\rho_{t+1}x_{t+1}\big|(d-\mu^\star)=\rho_{t}x_{t}\right)=1.
\end{align*}
\end{proposition}
\end{remark}

 We can conclude now that, for any pre-committed efficient mean-variance policy (except for the minimum variance policy), the probability that condition (i) or condition (ii) holds at time $t$ only depends on market parameters
$\P_i$ and $\K_i^\pm$, $i=0,1,\cdots, t-1$, where we assume the pre-committed mean-variance policy is efficient with $(d-\mu^\star)>\rho_0x_0$ (equivalent form of $d>\rho_0x_0$). This finding motivates us to deepen our analysis by linking the time consistency in efficiency with a minimum-variance signed supermartingale measure introduced in the next section.

\section{The variance-optimal signed supermartingale measure}

It has been well known that the problems of mean-variance
portfolio selection and mean-variance hedging have a strong
connection (see \cite{Schweizer2010}). \cite{XiaYan2006} showed
that in an unconstrained incomplete market, the optimal terminal
wealth of an efficient dynamic mean-variance policy is related to
the so-called variance-optimal signed martingale measure (VSMM) of
the market, and the optimal terminal wealth has a nonnegative
marginal utility if and only if VSMM is nonnegative. Note that
VSMM is the particular signed measure with the minimum variance among
all signed martingale measures, under which the discounted wealth
process of any admissible policy is a martingale. In discrete-time
unconstrained markets, the density of VSMM with respect to the
objective probability measure takes a product form (see
\cite{Schweizer1996} and \cite{CernyKallsen2009}). Actually, VSMM
plays a central role in the mean-variance hedging and is the
pricing kernel of the contingent claims (see \cite{Schweizer1995}
and \cite{Schweizer1996}).

Motivated by \cite{XiaYan2006}, we will carry out our analysis
forward in this section by deriving a similar ``VSMM'' in our constrained market.
However, the situation is much more complicated in a constrained
market than in an unconstrained one. \cite{PhamTouzi1999} and
\cite{FollmerSchied2004} showed that in a constrained market, no
arbitrage opportunity is equivalent to the existence of a
supermartingale measure, under which the discounted wealth process
of any admissible policy is a supermartingale (see
\cite{CarassusPhamTouzi} for a situation with upper bounds on
proportion positions). Therefore, we define in this paper the
particular measure with the minimum variance among all signed
supermartingale measures as the minimum-variance signed
supermartingale measure (VSSM) and derive its semi-analytical form
for discrete-time cone constrained markets. VSSM in our paper can
be considered as an extension of VSMM in constrained markets and
both take the product form. We will also show in this section
that the VSSM is not only related to the optimal terminal wealth
achieved by efficient mean-variance policies, but also associated
with TCIE of efficient mean-variance
policies. Our results explicitly assess the effect of portfolio
constraints on TCIE.

We use $\mathcal{L}^2(\mathcal{F}_{t+1},P)$ to denote the set of all $\mathcal{F}_{t+1}$-measurable square integrable random variables. According to \cite{PhamTouzi1999} and Chapter 9 of \cite{FollmerSchied2004}, a cone constrained market does not have any arbitrage opportunity if and only if there exists an equivalent probability measure under which the discounted wealth process of any admissible policy is supermartingale. Therefore, we extend the definitions of the signed martingale measure and the variance-optimal signed martingale measure proposed in \cite{Schweizer1996} to a signed supermartingale measure and minimum-variance signed supermartingale measure in this study.

\begin{definition}\label{def:SSM}
A signed measure $Q$ on $(\Omega,\mathcal{F}_T)$ is called a \emph{signed supermartingale measure} if $Q[\Omega]=1$, $Q \ll P$ with $dQ/dP\in \mathcal{L}^2(\mathcal{F}_T,P)$ and  the discounted wealth process of any admissible policy is supermartingale under $Q$, i.e., for $t=0,1,\cdots, T-1$,
\begin{align}
\mathbb{E}\left[\frac{dQ}{dP}\rho^{-1}_{t}x_T(\u_0,\u_1,\dots,\u_{T-1})\Big|\mathcal{F}_t\right]\leq x_t(\u_0,\u_1,\dots,\u_{t-1}),~~ \forall~ \u_i\in\mathcal{A}_i,\label{eq:supermartingale}
\end{align}
where $x_t(\u_0,\u_1,\dots,\u_{t-1})$ denotes the time-$t$ wealth level achieved by applying policy $\{\u_0,\u_1,\dots,\u_{t-1}\}$.
\end{definition}
We denote by $\mathbb{P}_s$ the set of all signed supermartingale measures. It is easy to see that inequality(\ref{eq:supermartingale}) is equivalent to either one of the following two inequalities,
\begin{align}
&\mathbb{E}\left[\frac{dQ}{dP}\P_t^\prime\u_t\Big|\mathcal{F}_t\right]\leq 0,~~ \forall~ \u_t\in\mathcal{A}_t,\label{eq:supermartingale-reduce-1}\\
&\mathbb{E}\left[\frac{dQ}{dP}\P_t\Big|\mathcal{F}_t\right]\in \mathcal{A}_t^\perp,\label{eq:supermartingale-reduce-2}
\end{align}
where $\mathcal{A}_{t}^\perp$ denotes the polar cone of $\mathcal{A}_{t}$, i.e.,
\begin{align*}
\mathcal{A}_{t}^\perp=\{\mathbf{y}\in\mathbb{R}^n~|~\mathbf{y}'\mathbf{x}\leq 0,~\mathbf{x}\in\mathcal{A}_{t}\}.
\end{align*}
\begin{definition}\label{def:VSSM}
A signed supermartingale measure $\tilde{P}$ is called \emph{minimum-variance signed supermartingale measure} if $\tilde{P}$ minimizes
\begin{align*}
{\rm Var}\left[\frac{dQ}{dP}\right]=\mathbb{E}\left[\left(\frac{dQ}{dP}-1\right)^2\right]=\mathbb{E}\left[\left(\frac{dQ}{dP}\right)^2\right]-1,
\end{align*}
over all $Q\in\mathbb{P}_s$.
\end{definition}

For $i=0,1,\cdots,T-1$, we define
\begin{align*}
m_i&=\mathbb{E}\left[\frac{d\tilde{P}}{dP}\Big|\mathcal{F}_{i}\right]\Big/\mathbb{E}\left[\frac{d\tilde{P}}{dP}\Big|\mathcal{F}_{i-1}\right].
\end{align*}
Then we have
\begin{align*}
\frac{d\tilde{P}}{dP}=m_1m_2\cdots m_{T}.
\end{align*}
If $m_i(\omega)=0$, we can set $m_j(\omega)$, $j>i$, equal to any value. It is easy to check that $\mathbb{E}[m_i|\mathcal{F}_{i-1}]=1$.

In the following, we will derive a semi-analytical form of the minimum-variance signed supermartingale measure in the cone constrained market. We first formulate the following pair of optimization problems for $t=0,1,\dots,T-1$,
$$
\begin{array}{rrl}
(A^+(t)):~~~& \min & \mathbb{E}\left[\left(\displaystyle\frac{1}{C_{t+1}^+}1_{\{m_{t+1}\geq 0\}}+\frac{1}{C_{t+1}^-}1_{\{m_{t+1}< 0\}}\right) m_{t+1}^2\Big|\mathcal{F}_{t}\right] \\ [3mm]
& \mbox{\rm s.t.} & \mathbb{E}\left[m_{t+1}|\mathcal{F}_{t}\right] = 1, \\ [2mm]
& & \mathbb{E}\big[m_{t+1}\P_{t}\Big|\mathcal{F}_{t}\big]\in\mathcal{A}_{t}^\perp,\\[2mm]
& & m_{t+1}\in\mathcal{L}^2(\mathcal{F}_{t+1},P),
\end{array}
$$
and
$$
\begin{array}{rrl}
(A^-(t)):~~~& \min & \mathbb{E}\left[\left(\displaystyle\frac{1}{C_{t+1}^+}1_{\{m_{t+1}\leq 0\}}+\frac{1}{C_{t+1}^-}1_{\{m_{t+1}> 0\}}\right) m_{t+1}^2\big|\mathcal{F}_{t}\right] \\ [3mm]
& \mbox{\rm s.t.} & \mathbb{E}\left[m_{t+1}|\mathcal{F}_{t}\right] = 1, \\ [2mm]
& & -\mathbb{E}\big[m_{t+1}\P_{t}\big|\mathcal{F}_{t}\big]\in\mathcal{A}_{t}^\perp,\\[2mm]
& & m_{t+1}\in\mathcal{L}^2(\mathcal{F}_{t+1},P).
\end{array}
$$

\begin{lemma}\label{lem:vssm} \sl

The solutions of $(A^+(t))$ and $(A^-(t))$ are given respectively by
\begin{align*}
m_{t+1}^+&=\frac{1}{C_{t}^+}\left[C_{t+1}^+(1-\P_t^\prime\K_t^{+})1_{\{m_{t+1}^+\geq 0\}}+C_{t+1}^-(1-\P_t^\prime\K_t^{+})1_{\{m_{t+1}^+< 0\}}\right],\\
m_{t+1}^-&=\frac{1}{C_{t}^-}\left[C_{t+1}^+(1+\P_t^\prime\K_t^{-})1_{\{m_{t+1}^-\leq 0\}}+C_{t+1}^-(1+\P_t^\prime\K_t^{-})1_{\{m_{t+1}^-> 0\}}\right],
\end{align*}
and the optimal objective values of $(A^+(t))$ and $(A^-(t))$ are $\frac{1}{C_t^+}$ and $\frac{1}{C_t^-}$ respectively.
\end{lemma}

\begin{theorem}\label{prop:VSSM} \sl
The density of the minimum-variance signed supermartingale measure $\tilde{P}$ (with respect to objective probability measure $P$) is given by
\begin{align*}
\frac{d\tilde{P}}{dP}&=(C_0^+)^{-1}\prod_{i=0}^{T-1}B_i,
\end{align*}
where
\begin{align*}
B_0&=1-\P_0^\prime\K_0^+,\\
B_i&=(1-\P_i^\prime\K_i^+)1_{\{\prod_{j=0}^{i-1}B_j\geq 0\}}+(1+\P_i^\prime\K_i^-)1_{\{\prod_{j=0}^{i-1}B_j<0\}},~i=1,2,\cdots,T-1.
\end{align*}
Furthermore,
\begin{align}
&\mathbb{E}\left[\frac{d\tilde{P}}{dP}\Big|\mathcal{F}_t\right]=(C_0^+)^{-1}\prod_{i=0}^{t-1}B_i\left(C_t^+1_{\{\prod_{j=0}^{t-1}B_j\geq 0\}}+C_t^-1_{\{\prod_{j=0}^{t-1}B_j< 0\}}\right),\label{eq:conditional-expectation-VSSM}\\
&\mathbb{E}\left[\left(\frac{d\tilde{P}}{dP}\right)^2\right]=\frac{1}{C_0^+}.\label{eq:second-monent-VSSM}
\end{align}
\end{theorem}

There is a strong connection between VSSM and the optimal terminal wealth achieved by the pre-committed efficient mean-variance policy. Substituting the pre-committed efficient mean-variance policy in (\ref{eq:opt-u-star}) into the wealth dynamic equation yields
\begin{equation}\label{eq:xstar}
x_{t+1}^\star=\left\{\begin{array}{ll}
    s_tx_t^\star + s_t\P_t'\K_t^+\big((d-\mu^{\star})\rho_t^{-1} - x_t^\star\big), & \mbox{if}~~d-\mu^{\star}\geq \rho_t x_{t}^\star, \\ [2mm]
    s_tx_t^\star - s_t\P_t'\K_t^-\big((d-\mu^{\star})\rho_t^{-1} - x_t^\star\big), & \mbox{if}~~d-\mu^{\star}< \rho_t x_{t}^\star,
  \end{array}\right.
\end{equation}
with $x_0^\star=x_0$. Set $y_t^\star \triangleq x_t^\star-(d-\mu^{\star})\rho_t^{-1}$.
From the wealth equation in (\ref{eq:xstar}) which $x_t^\star$
satisfies, we deduce
\begin{equation}\label{eq:y-system-discrete}
\left\{\begin{array}{l}
y_{t+1}^\star = s_t y_t^\star - s_t\P_t'\K_t^+y_t^\star1_{\{ y_t^\star\leq 0\}}+s_t\P_t'\K_t^-y_t^\star1_{\{ y_t^\star> 0\}}, \\ [2mm]
y_0^\star= x_0-(d - \mu^\star)\rho_0^{-1}.
\end{array}\right.
\end{equation}
Note that $y_0^\star=x_0-(d - \mu^\star)\rho_0^{-1} = \frac{d\rho_0^{-1}-x_0}{C_0^+-1} \leq 0$ by virtue of the fact that
$d \geq x_0\rho_0$ and $C_0^+<1$.

We can show
\begin{align*}
y_t^\star = y_0^\star\displaystyle\prod_{i=0}^{t-1}s_i\prod_{i=0}^{t-1}B_i, \quad t = 1, 2, \cdots, T.
\end{align*}
For $t=1$, it is trivial. Assume that the statement holds true for $t$, we now show that the statement also holds true for $t+1$, as
\begin{align*}
y_{t+1}^\star&=s_t y_t^\star - s_t\P_t'\K_t^+y_t^\star1_{\{ y_t^\star\leq 0\}}+s_t\P_t'\K_t^-y_t^\star1_{\{ y_t^\star> 0\}}\\
&=y_0^\star\displaystyle\prod_{i=0}^{t}s_i\prod_{i=0}^{t-1}B_i \left[(1-\P_t'\K_t^+)1_{\{\prod_{j=0}^{t-1}B_j\geq 0\}}+(1+\P_t'\K_t^-)1_{\{\prod_{j=0}^{t-1}B_j< 0\}}\right].
\end{align*}
Thus, the time $t$ optimal wealth achieved by the pre-committed efficient mean-variance policy
is given by
\begin{equation}\label{eq:x-star-discrete}
x_t^\star = (d-\mu^\star)\rho_t^{-1} - [(d - \mu^\star) - x_0\rho_0]\rho_t^{-1}\prod_{i=0}^{t-1}B_i,
\end{equation}
which leads to the following theorem.

\begin{theorem} \label{thm:wealth-VSSM}\sl
The optimal terminal wealth achieved by the pre-committed efficient mean-variance policy $x_T^\star$ and the VSSM $\tilde{P}$ have the following duality relationship:
\begin{align*}
x_T^\star = (d-\mu^\star) - \frac{(d - \mu^\star) - x_0\rho_0}{\mathbb{E}\left[\left(\frac{d\tilde{P}}{dP}\right)^2\right]}\cdot\frac{d\tilde{P}}{dP}.
\end{align*}
\end{theorem}

\begin{remark}
\cite{XiaYan2006} considered the mean-variance portfolio selection problem in an incomplete, albeit unconstrained, market and established the relationship between the mean-variance efficient portfolio and the variance-optimal signed martingale measure (VSMM) via analyzing the geometric property of the problem. Actually, the above Theorem \ref{thm:wealth-VSSM} is an extension of Theorem 3.1 in \cite{XiaYan2006} for the discrete-time cone constrained market. When the convex cone constraint is chosen as the whole space, our theorem reduces to the result in \cite{XiaYan2006}. On the other hand, different from \cite{XiaYan2006}, we prove the theorem by solving both the optimal terminal wealth and the VSSM directly.
\end{remark}

Most prominently, we will demonstrate in the following that the VSSM is also related to the property of TCIE of the pre-committed efficient mean-variance policy.

\begin{theorem}\label{thm:TCIE-VSSM} \sl
The pre-committed efficient mean-variance policy (except for the minimum variance policy) in a cone constrained market is TCIE if and only if the variance-optimal signed supermartingale measure of this market satisfies:
\begin{align}\label{eq:condition-1}
~\mathbb{E}\left[\frac{d\tilde{P}}{dP}\Big|\mathcal{F}_t\right](\omega)\geq 0, ~\forall~ 0<t<T,~\forall~\omega\in\Omega;
\end{align}
or
\begin{align}\label{eq:condition-2}
~\mathbb{E}\left[\frac{d\tilde{P}}{dP}\Big|\mathcal{F}_k\right](\omega)=\mathbb{E}\left[\frac{d\tilde{P}}{dP}\Big|\mathcal{F}_{\tau}\right](\omega)< 0,~\forall~\tau\leq k\leq T,~\forall~\omega\in\Omega,~
\end{align}
where the stopping time $\tau$ is defined as
\begin{align*}
\tau=\inf\left\{~t~ \Big\vert ~\mathbb{E}\left[\frac{d\tilde{P}}{dP}\Big|\mathcal{F}_t\right]< 0, ~~t = 1, 2, \cdots, T\right\}.
\end{align*}
\end{theorem}


We can conclude from Theorem \ref{thm:TCIE-VSSM}
 that the pre-committed efficient mean-variance policy (except for the minimum variance policy) satisfies TCIE if and only if the conditional expectation of VSSM's density (respect to the original probability measure) is nonnegative, or once the conditional expectation takes a negative value, it remains the same value until the terminal time.

It is also easy to see that condition (\ref{eq:condition-1}) implies that $\forall~\omega\in\Omega,~(1-\P_t^\prime(\omega)\K_t^+)\geq 0$ and
\begin{align*}
\u_t^\star=s_t\K_t^+\big((d-\mu^{\star})\rho_t^{-1}-x_t\big).
\end{align*}
In such a case, every mean-variance investor holds a long position of the effcieint risky portfolio $\K_t^+$, whose excess rate of return does not exceed 100\% ($\P_t^\prime\K_t^+\leq 1$), and achieves efficiency during the entire investment horizon.

The stopping time $\tau$ can be also expressed as
\begin{align*}
\tau=\inf\left\{~t~ \big\vert ~(1-\P_{t-1}^\prime\K_{t-1}^+)< 0, ~~t = 1, 2, \cdots, T\right\}.
\end{align*}
Then, condition (\ref{eq:condition-2}) implies that for $t<\tau$,
\begin{align*}
\u_t^\star=s_t\K_t^+\big((d-\mu^{\star})\rho_t^{-1}-x_t\big),
\end{align*}
and for $k\geq \tau$,
\begin{align*}
\K_k^-=\0,~~\u_k^\star=s_k\K_k^-\big((d-\mu^{\star})\rho_k^{-1}-x_k\big)=\0.
\end{align*}
In this situation, every mean-variance investor starts from holding a long position of the efficient risky portfolio $\K_t^+$ and switches all his wealth into the  riskless asset once the excess rate of return of $\K_t^+$ exceeds 100\%, i.e., $\P_t^\prime\K_t^+>1$.

Theorem \ref{thm:TCIE-VSSM} shows that whether the pre-committed efficient mean-variance policy (except for the minimum variance policy) is TCIE only depends on the basic market setting (the distribution of excess rate of return $\P_t$ and the portfolio constraint set $\mathcal{A}_t$) and does not depend on the initial wealth level, $x_0$, and the objective level which the investor aspires to achieve, $d$. This clear recognition motivates us to consider active introduction of additional market constraints such that the phenomenon of time inconsistency in efficiency can be eliminated.

\section{Elimination of time inconsistency in efficiency with portfolio constraints}
From our discussion in the previous sections, it becomes clear that constraints on portfolio do have effects on TCIE. Suppose that a given discrete-time mean-variance problem is originally not TCIE. Are we able to eliminate the time inconsistency in efficiency by introducing suitable portfolio constraints into the market? We will demonstrate a positive answer to this question in this section.

\begin{remark}
We proceed our investigation starting from an unconstrained market, then a market with no shorting, before dealing with a general cone constrained market.

\bigskip

i) Case of unconstrained markets: \\

If the market is constraint free, i.e., $\mathcal{A}_t=\mathbb{R}^n$, we have
\begin{align*}
\K_t^\pm&=\pm\E^{-1}\left[\P_t\P_t^{\prime}\right]\E\left[\P_t\right],\\
C_t^\pm&=\prod_{i=t}^{T-1}(1-\E\left[\P_i^\prime\right]\E^{-1}\left[\P_i\P_i^{\prime}\right]\E\left[\P_i\right]).
\end{align*}
Therefore, the optimal mean-variance policy of $(P(d))$ is
\begin{align*}
\u_t^\star=s_t\big((d-\mu^{\star})\rho_t^{-1}-x_t\big)\E^{-1}\left[\P_t\P_t^{\prime}\right]\E\left[\P_t\right], ~~t=0,1,\dots,T-1,
\end{align*}
where
\begin{align}\label{mu_unconstrained}
\mu^{\star} &= \displaystyle\frac{d-\rho_0 x_{0}}{1-\prod_{i=0}^{T-1}(1-\E\left[\P_i^\prime\right]\E^{-1}\left[\P_i\P_i^{\prime}\right]\E\left[\P_i\right])^{-1}},
\end{align}
which is exactly the result in \cite{LN}. We can assume here that $\E^{-1}\left[\P_t\P_t^{\prime}\right]\E\left[\P_t\right]\neq\0$. Otherwise, all efficient policies reduce to the one corresponding to investing only in the riskless asset.

Furthermore, the minimum-variance signed supermartingale measure in the unconstrained market is given by
\begin{align*}
\frac{d\tilde{P}}{dP}=\prod_{i=0}^{T-1}\frac{1-\P_i^\prime\E^{-1}\left[\P_i\P_i^{\prime}\right]\E\left[\P_i\right]}{1-\E\left[\P_i^\prime\right]\E^{-1}\left[\P_i\P_i^{\prime}\right]\E\left[\P_i\right]},
\end{align*}
which is exactly the variance-optimal signed martingale measure (VSMM) obtained in \cite{Schweizer1995}, \cite{Schweizer1996} and \cite{CernyKallsen2009}.

Theorem \ref{thm:TCIE-VSSM} shows that the pre-committed efficient mean-variance policy (except for the minimum variance policy) in the unconstrained market satisfies time consistency in efficiency if and only if VSMM is a nonnegative measure for any $\mathcal{F}_t$, i.e.,
\begin{align}\label{eqn:TCIE-unconstrain}
\P_i^\prime\E^{-1}\left[\P_i\P_i^{\prime}\right]\E\left[\P_i\right]\leq 1,~~\mbox{a.s.}
\end{align}
Actually, \cite{CuiLiWangZhu2012} proved that condition (\ref{eqn:TCIE-unconstrain}) does not hold only if the market is an incomplete market and proposed a TCIE revised policy which i) achieves the same mean-variance pair as the pre-committed efficient policy does and ii) receives an additional positive free cash flow during the investment horizon.

\bigskip

ii) Case of markets without shorting:\\

Assume that shorting of risky assets is not allowed in the market, i.e., $\mathcal{A}_t=\mathbb{R}_+^n$, and the expected excess rate of return of risky assets is nonnegative, i.e., $\E[\P_t] \geq \0$. In this situation, we have
\begin{align*}
\K_t^+&=\arg\min_{\K_t\in\mathbb{R}_+^n}\E\left[C_{t+1}^+\Big( 1-\P_t^{\prime}\K_t\Big)^21_{\{\P_t^{\prime}\K_t\leq 1\}}+C_{t+1}^-\Big( 1-\P_t^{\prime}\K_t\Big)^21_{\{\P_t^{\prime}\K_t> 1\}}\right], \\
\K_t^-&=\arg\min_{\K_t\in\mathbb{R}_+^n}\E\left[C_{t+1}^+\Big( 1+\P_t^{\prime}\K_t\Big)^21_{\{\P_t^{\prime}\K_t\leq -1\}}+C_{t+1}^-\Big( 1+\P_t^{\prime}\K_t\Big)^21_{\{\P_t^{\prime}\K_t> -1\}}\right]=\0.
\end{align*}
In addition, we also have
\begin{align*}
\left(\bigtriangledown_{\K_t} h_t^{-}(\0)\right)^\prime(\K_t-\0)=2C_{t+1}^-\E[\P_t^\prime]\K_t\geq 0,~\forall~ \K_t\in\mathbb{R}_+^n.
\end{align*}
Therefore, the optimal policy of $(P(d))$ is
\begin{align}
\u_t^\star=s_t\big((d-\mu^{\star})\rho_t^{-1}-x_t\big)\K_t^+1_{\{d-\mu^{\star}\geq \rho_tx_t\}}, ~~t=0,1,\dots,T-1,
\end{align}
where
\begin{align*}
\mu^{\star} &= \displaystyle\frac{d-\rho_0 x_{0}}{1-(C_0^+)^{-1}},
\end{align*}
which is the result derived in \cite{CuiGaoLiLi2012}.

Furthermore, the variance-optimal signed supermartingale measure in such a market setting is given by
\begin{align*}
\frac{d\tilde{P}}{dP}&=(C_0^+)^{-1}\prod_{i=0}^{(T-1)\wedge(\tau-1)}(1-\P_i^\prime\K_i^+),
\end{align*}
where
\begin{align*}
\tau=\inf\left\{~t~ \big\vert ~(1-\P_{t-1}^\prime\K_{t-1}^+)< 0, ~~t = 1, 2, \cdots, T\right\}.
\end{align*}
We can see that $C_t^-=1$, $t=0,1,\cdots,T-1$. Therefore, according to Theorem \ref{thm:TCIE-VSSM}, all pre-committed efficient policies are TCIE in a market with no shorting and with nonnegative expected excess rate of return.
\end{remark}

We proceed now to a discussion for a general cone-constrained market setting.

\begin{theorem} \label{prop:condition-TCIE}\sl
If a convex cone $\mathcal{A}_t$ is chosen to restrict portfolios such that the expected excess rate of return vector $\E[\P_t]$ lies in the dual cone of $\mathcal{A}_t$, i.e.,
\begin{align*}
\E[\P_t]\in\mathcal{A}_t^*,
\end{align*}
where $\mathcal{A}_t^*=\{\mathbf{y}\in\mathbb{R}^n~|~\mathbf{y}^\prime\mathbf{x}\geq 0,~\mathbf{x}\in\mathcal{A}_t\}=-\mathcal{A}_t^\perp$, then the corresponding optimal discrete-time pre-committed efficient mean-variance policy is TCIE.
\end{theorem}

\begin{figure}[ht]
  \centering
  \includegraphics[width=10cm]{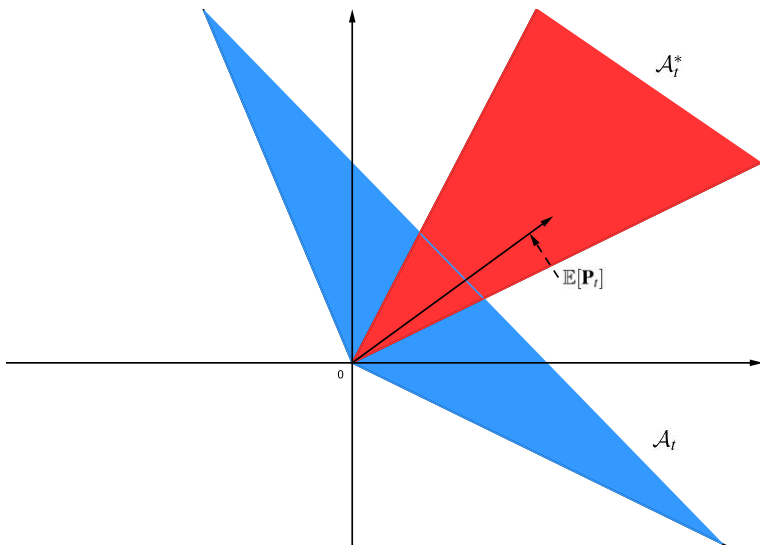}
  \label{dual-cone}
  \caption{Construction of Suitable Cone Constraint}
\end{figure}

Figure \ref{dual-cone} illustrates the above proposition graphically. Basically, this is an inverse process to find the convex cone $\mathcal{A}_t$. For a given market, $\E[\P_t]$ is known. We first identify a cone $\mathcal{A}_t^*$ such that $\E[\P_t]\in\mathcal{A}_t^*$. We then find another cone $\mathcal{A}_t$ such that the selected $\mathcal{A}_t^*$ becomes its dual cone. Apparently, the condition in Theorem \ref{prop:condition-TCIE} aims to enforce the inefficient risky portfolio $\K_t^-$ equal to zero in order to achieve condition (\ref{eq:condition-2}). Note that condition (\ref{eq:condition-1}) is much harder to satisfy, as it is related to the distribution of excess rate of return which is uncontrollable in general.


\begin{exn}\label{ex3} \rm
We now consider an example of constructing a three-year pension fund consisting of S\&P 500 (SP), the index of Emerging Market (EM), Small Stock (MS) of U.S market and a bank account. The annual rates of return of these three indices have the expected values, variances and correlations given in Table \ref{Table1}, based on the data provided in \cite{Elton:2007}.

\begin{table}[h]
  \centering
  \begin{tabular}{lccc}
    \hline
                    & SP       & EM      & MS \\
    \hline
    Expected Return & $14\%$   & $16\% $ & $17\%$ \\
    Variance        & $18.5\%$ & $30\% $ & $24\%$ \\
    \hline
    \multicolumn{4}{c}{Correlation}\\
    \hline
    SP              & $1$      & $0.64$     & $0.79$ \\
    EM              &          & $1$        & $0.75$ \\
    MS              &          &            & $1$ \\
    \hline
  \end{tabular}
  \caption{Data for Example \ref{ex3}}\label{Table1}
\end{table}

We further assume that all annual rates of return are statistically
independent and follow i) the identical multivariate normal distribution (with the statistics described above) or ii) the identical multivariate $t$ distribution with freedom 5 (and with the statistics described above) for all $3$ years, and the
annual risk free rate is $5\%$, i.e., $s_t=1.05$, $t=0,1,2$. We
first compute  $\E[\P_t]$, $\mbox{\rm Cov}(\P_t)$ and $\E[\P_t\P_t^\prime]$ as follows,
for $t=0,1,2$,
\begin{align}
\E[\P_t]=\begin{bmatrix}
             0.09\\
             0.11\\
             0.12\\
         \end{bmatrix},
~
\mbox{\rm Cov}(\P_t)=\begin{bmatrix}
    0.0342  &  0.0355  &  0.0351\\
    0.0355  &  0.0900  &  0.0540\\
    0.0351  &  0.0540  &  0.0576\\
                                               \end{bmatrix},
~\E[\P_t\P_t^\prime]=\begin{bmatrix}
    0.0423  &  0.0454  &  0.0459 \\
    0.0454  &  0.1021  &  0.0672 \\
    0.0459  &  0.0672  &  0.0720 \\
                                               \end{bmatrix}.
\end{align}
 In order to examine the phenomenon of time inconsistency in efficiency (by observing the number that the wealth level exceeds the threshold $(d-\mu^\star)\rho_t^{-1}$), we simulate $2\times10^6$ samples paths for each distribution assumption, with the setting of initial wealth equal to $x_0=1$ and the target expected return equal to $d=1.35$.

\textbf{Case 1}: When the market is unconstrained, the optimal mean-variance policy of $(P(d))$ is
\begin{align*}
\u_t^\star=s_t\big((d-\mu^{\star})\rho_t^{-1}-x_t\big)\E^{-1}\left[\P_t\P_t^{\prime}\right]\E\left[\P_t\right]
=1.05\left((1.35+0.1808)1.05^{t-3}-x_t\right)\begin{bmatrix}
    1.0580\\
   -0.1207\\
    1.1052\\
         \end{bmatrix},\\
~~t=0,1,2,
\end{align*}
with $\mu^\star=-0.1808$ (based on (\ref{mu_unconstrained})) for both distribution assumptions.  Apparently, under both the {\it unbounded} multivariate normal distribution and multivariate $t$ distribution, equation (\ref{eqn:TCIE-unconstrain}) does not hold, which implies that the time inconsistency in efficiency may occur. More specifically, recalling Theorem \ref{prop:truncated-efficency} and Lemma \ref{lem:TCIE-condition} and noticing $C_t^+=C_t^-<1$ with $t<T$, the pre-committed efficient mean-variance policy does not satisfy TCIE if and only if the optimal wealth level $x_t^\star$ exceeds the threshold
\begin{align*}
(d-\mu^\star)\rho_t^{-1}=(1.35+0.1808)*1.05^{t-3}.
\end{align*}
The simulation results show that the probabilities that $x_t^\star$ exceeds the threshold $(d-\mu^\star)\rho_t^{-1}$ are 0.055 for the multivariate normal distribution and 0.0558 for the multivariate $t$ distribution. This simulation outcome indicates that a distribution with a heavier tail tends to demonstrate a higher degree of time inconsistency in efficiency in an unconstrained market.

\textbf{Case 2}: To eliminate the time inconsistency in efficiency, we consider first to add the following cone constraint to the market,
\begin{align*}
\mathcal{A}_t=\{\u_t\in\mathbb{R}^n~|~\E[\P_t^\prime]\u_t\geq 0\},
\end{align*}
which is a half-space with boundary $\E[\P_t^\prime]\u_t= 0$ that is a hyperplane orthogonal to $\E[\P_t]$. The dual cone of $\mathcal{A}_t$ is
\begin{align*}
\mathcal{A}_t^*=\{\mathbf{y}\in\mathbb{R}^n~|~\mathbf{y}=\lambda\E[\P_t], ~\lambda\geq 0\},
\end{align*}
which is exactly the ray along $\E[\P_t]$ (see Proposition 3.2.1 of \cite{Bertsekas2003}). Notice that the constraint cone, $\mathcal{A}_t$, defined above is the largest cone (thus the loosest constraint) which we can identify to eliminate the time inconsistency in efficiency in this example.

Based on the proof of Theorem \ref{prop:condition-TCIE}, we have $\K_0^{-}=\K_1^{-}=\K_2^{-}=\0$ for both distribution assumptions. By Lemma \ref{lem:K-C}, we can compute $\K_t^{+}$ numerically through penalty function method (see Appendix A of \cite{CuiGaoLiLi2012}) with initial point $[1.06, -0.12, 1.11]'$ as
\begin{align*}
\mbox{i)}~~ &\K_0^{+}=\begin{bmatrix}
    1.0589\\
   -0.1212\\
    1.1086\\
     \end{bmatrix},
~\K_1^{+}=\begin{bmatrix}
    1.0600\\
   -0.1200\\
    1.1100\\
     \end{bmatrix},
~\K_2^{+}=\begin{bmatrix}
    1.0600\\
   -0.1200\\
    1.1100\\
     \end{bmatrix}
\end{align*}
for the multivariate normal distribution and
\begin{align*}
\mbox{ii)}~~ &\K_0^{+}=\begin{bmatrix}
    1.0461\\
   -0.1335\\
    1.0929\\
     \end{bmatrix},
~\K_1^{+}=\begin{bmatrix}
    1.0548\\
   -0.1263\\
    1.1034\\
     \end{bmatrix},
~\K_2^{+}=\begin{bmatrix}
    1.0600\\
   -0.1200\\
    1.1100\\
     \end{bmatrix}
\end{align*}
for the multivariate $t$ distribution.
The optimal investment policy is thus
\begin{align*}
\mbox{i)}~~&\u_t^\star(x_t)=1.05\big((1.35+0.1810)1.05^{t-3}-x_t\big)\K_t^{+}1_{\{x_t<1.05^{(t-3)}(1.5310)\}}
\end{align*}
for the multivariate normal distribution and
\begin{align*}
\mbox{ii)}~~&\u_t^\star(x_t)=1.05\big((1.35+0.1831)1.05^{t-3}-x_t\big)\K_t^{+}1_{\{x_t<1.05^{(t-3)}(1.5331)\}}
\end{align*}
for the multivariate $t$ distribution. The simulation shows that the probabilities that $x_t^\star$ exceeds the threshold $(d-\mu^\star)\rho_t^{-1}$ are 0.0559 for the multivariate $t$ distribution and 0.0533 for the multivariate $t$ distribution. Once $x_t^\star$ exceeds the threshold $(d-\mu^\star)\rho_t^{-1}$, the investor puts all his wealth into the riskless asset, which eliminates the time inconsistency in efficiency in this example.

\textbf{Case 3}: In this case, we introduce into the market a more realistic convex cone constraint,
\begin{align*}
\mathcal{A}_t=\{\u_t\in\mathbb{R}^n~|~u_t^2\geq 0,~u_t^3\geq 0,~u_t^1+u_t^2+u_t^3\geq 0\},
\end{align*}
which implies that short selling is not allowed for the index of Emerging Market and the Small Stock of U.S market, and the negative position on S\&P 500 cannot be too large. The dual cone of $\mathcal{A}_t$ in this case is
\begin{align*}
\mathcal{A}_t^*=\left\{\mathbf{y}\in\mathbb{R}^n~\Bigg|~\mathbf{y}=\begin{bmatrix}
                                                                     1 & 0 & 0 \\
                                                                     1 & 1 & 0 \\
                                                                     1 & 0 & 1 \\
                                                                   \end{bmatrix}\lambda, ~\lambda\geq \0\right\}.
\end{align*}
Note that specifying $\lambda$ at $[0.09,0.02,0.03]'$ $\in$ $\mathcal{A}_t^*$ yields the ray along $\E[\P_t]$.

Based on the proof in Theorem \ref{prop:condition-TCIE}, we have $\K_0^{-}=\K_1^{-}=\K_2^{-}=\0$ for both distribution assumptions. By Lemma \ref{lem:K-C}, we can compute $\K_t^{+}$ numerically through penalty function method (see Appendix A of \cite{CuiGaoLiLi2012}) with initial point $[1.06, 0.05, 1.11]'$ as
\begin{align*}
\mbox{i)}~~ &\K_0^{+}=\begin{bmatrix}
    1.0076\\
    0.0044\\
    1.0324\\
     \end{bmatrix},
~\K_1^{+}=\begin{bmatrix}
    1.0133\\
    0.0037\\
    1.0373\\
     \end{bmatrix},
~\K_2^{+}=\begin{bmatrix}
    1.0147\\
    0.0031\\
    1.0401\\
     \end{bmatrix}
\end{align*}
for the multivariate normal distribution and
\begin{align*}
\mbox{ii)}~~ &\K_0^{+}=\begin{bmatrix}
    1.0112\\
    0.0030\\
    1.0413\\
     \end{bmatrix},
~\K_1^{+}=\begin{bmatrix}
    1.0201\\
    0.0026\\
    1.0522\\
     \end{bmatrix},
~\K_2^{+}=\begin{bmatrix}
    1.0216\\
    0.0039\\
    1.0501\\
     \end{bmatrix}
\end{align*}
for the multivariate $t$ distribution. The optimal investment policy is thus
\begin{align*}
\mbox{i)}~~&\u_t^\star(x_t)=1.05\big((1.35+0.1818)1.05^{t-3}-x_t\big)\K_t^{+}1_{\{x_t<1.05^{(t-3)}(1.5318)\}}
\end{align*}
for the multivariate normal distribution and
\begin{align*}
\mbox{ii)}~~&\u_t^\star(x_t)=1.05\big((1.35+0.1843)1.05^{t-3}-x_t\big)\K_t^{+}1_{\{x_t<1.05^{(t-3)}(1.5343)\}}
\end{align*}
for the multivariate $t$ distribution. The simulation shows that the probabilities that $x_t^\star$ exceeds the threshold $(d-\mu^\star)\rho_t^{-1}$ are 0.0569 for the multivariate normal distribution and 0.0588 for the multivariate $t$ distribution and. Although, compared to the unconstrained case, both the probabilities increase, the investor puts all his wealth into the riskless asset immediately after $x_t^\star$ exceeds the threshold $(d-\mu^\star)\rho_t^{-1}$.

 For the unconstrained market in Case 1, the expression of the efficient frontier achieved by the pre-committed policy  is given in (76) in \cite{LN}. For cone constrained markets in Case 2 and Case 3, their efficient frontiers achieved by the pre-committed policy are given in Theorem \ref{prop:cone-constained} of this paper. For problem $(P(d))$, we also derive in Appendix A9 of this paper its efficient frontier achieved by the time consistent policy proposed by \cite{BasakChabakauri} and \cite{Bjork:2012}, with its expression given in (\ref{eq:eff-frontier-TC}).

Figure \ref{fig:eff-frontiers-1} depicts  the efficient frontiers in the mean-standard deviation space for Case 1, Case 2 and Case 3 and demonstrates a clear domination relationship among the three. Furthermore, Figure \ref{fig:eff-frontiers-2} illustrates  a clear dominance relationship between Case 3 and the efficient frontier achieved by the time consistent policy. As both TCIE policies and the time consistent policy aim to align the inherently inconsistent global and local interests, they all sacrifice certain degrees of global performance, thus all being dominated by the pre-committed policy. Case 2 dominates Case 3 as Case 2 is associated with a looser constraint, while Case 3 is associated with a tighter constraint. It is interesting to note that both TCIE policies dominate the time consistent policy significantly, which indicates that insisting time consistency for an inherently time inconsistent problem may suffer a significant loss in its global performance. Expression (\ref{eq:eff-frontier-TC}) reveals that the time consistent policy achieves a good efficient frontier globally only if $B_t$ is large. In conclusion, by introducing appropriate constraints into the model, we can not only eliminate time inconsistency in efficiency, but also strick a good balance between the global and local mean-variance efficiency. Actually, relaxing the time consistency requirement to TCIE offers us a flexibility in deciding which level of a good global performance to maintain by introducing suitable portfolio constraints and deriving the corresponding pre-committed TCIE policy.

\begin{figure}[h]
\subfigure[Multivariate $t$ distribution]{
\begin{minipage}[t]{0.5\linewidth}
   \includegraphics[width=8.5cm]{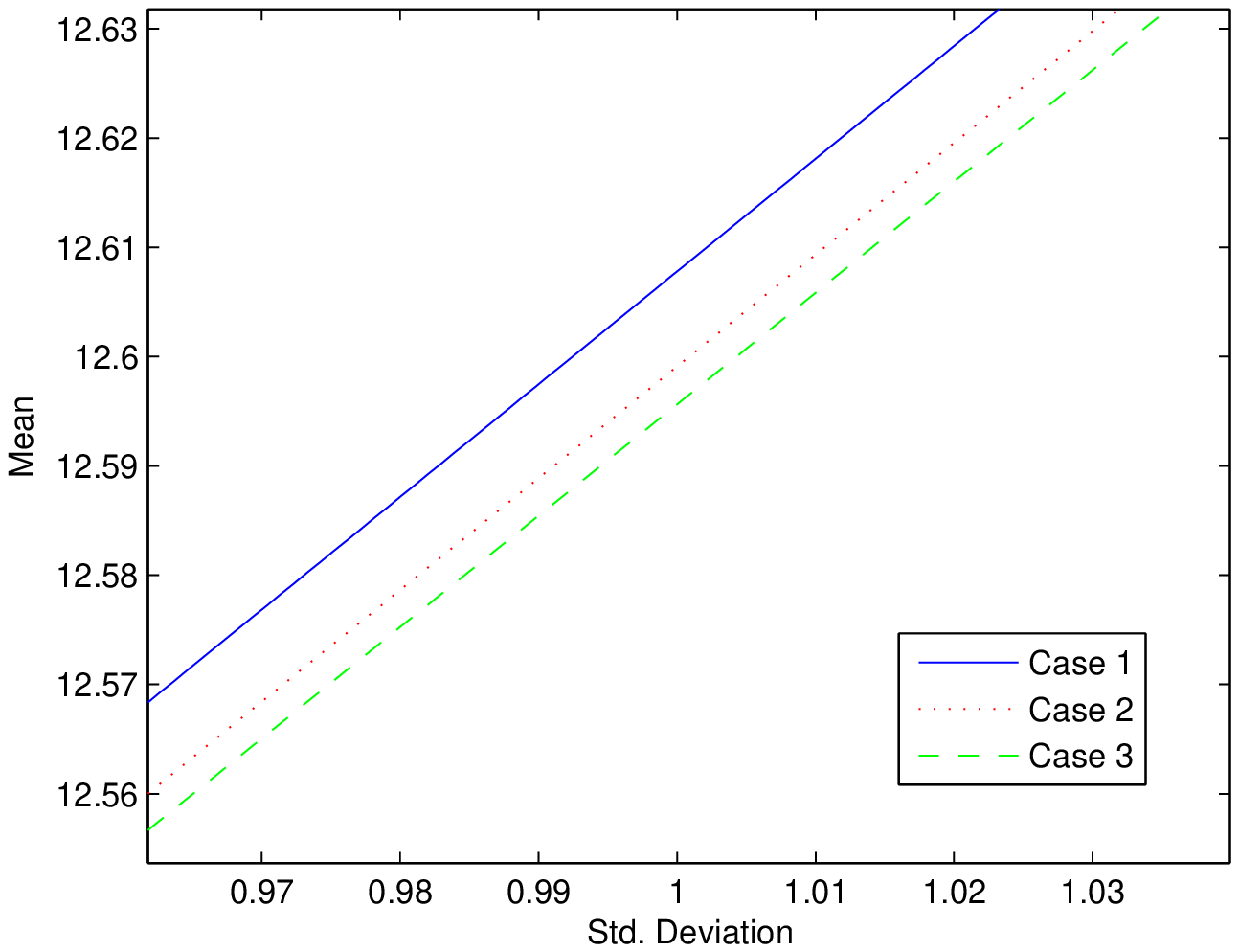}
   \end{minipage}}%
\subfigure[Multivariate normal distribution]{
\begin{minipage}[t]{0.5\linewidth}
   \includegraphics[width=8.5cm]{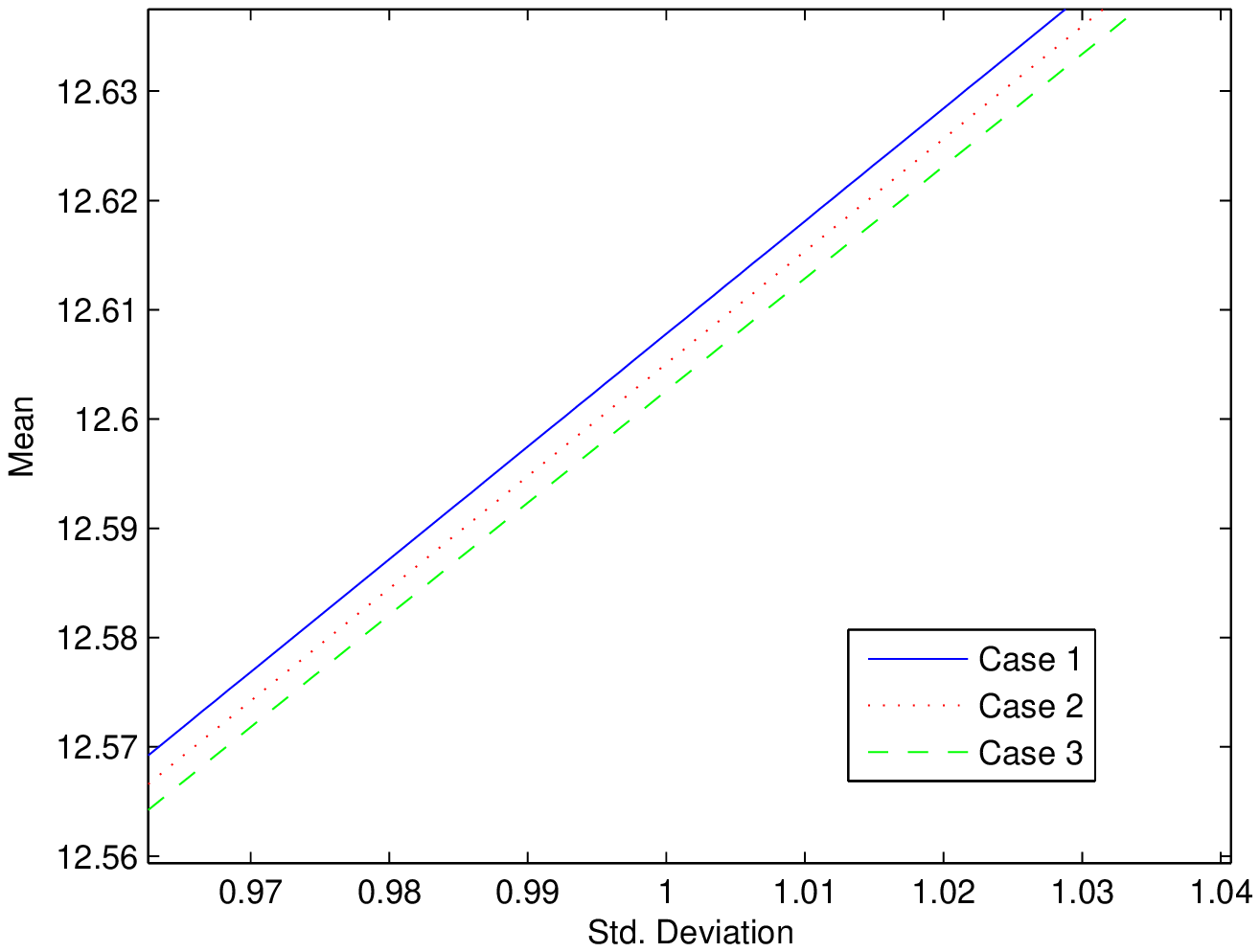}
   \end{minipage}}%
\caption{Comparison of efficient frontiers for Cases 1, 2 and 3 of Example \ref{ex3}} \label{fig:eff-frontiers-1}
\end{figure}

\begin{figure}[h]
\subfigure[Multivariate $t$ distribution]{
\begin{minipage}[t]{0.5\linewidth}
   \includegraphics[width=8.5cm]{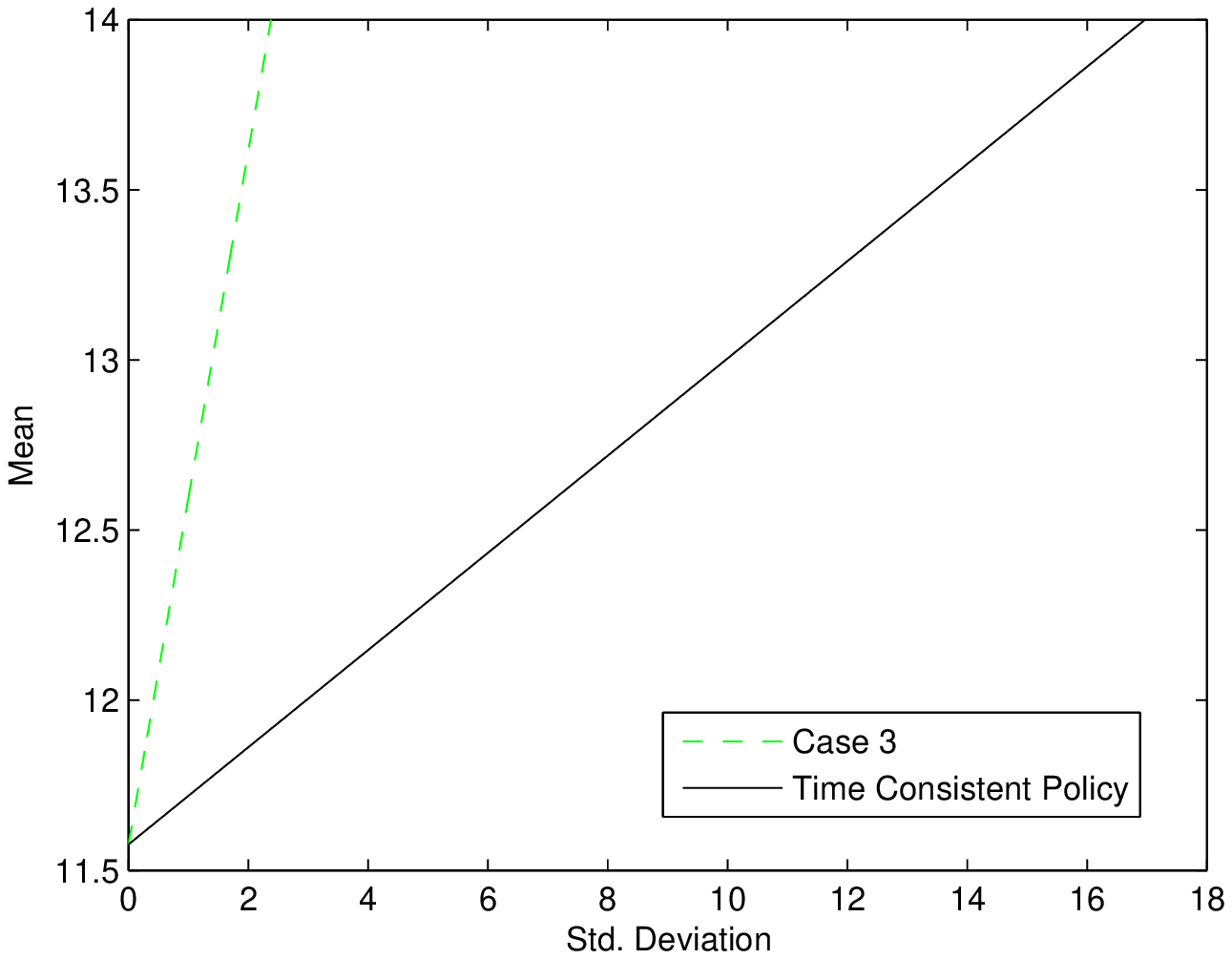}
   \end{minipage}}%
\subfigure[Multivariate normal distribution]{
\begin{minipage}[t]{0.5\linewidth}
   \includegraphics[width=8.5cm]{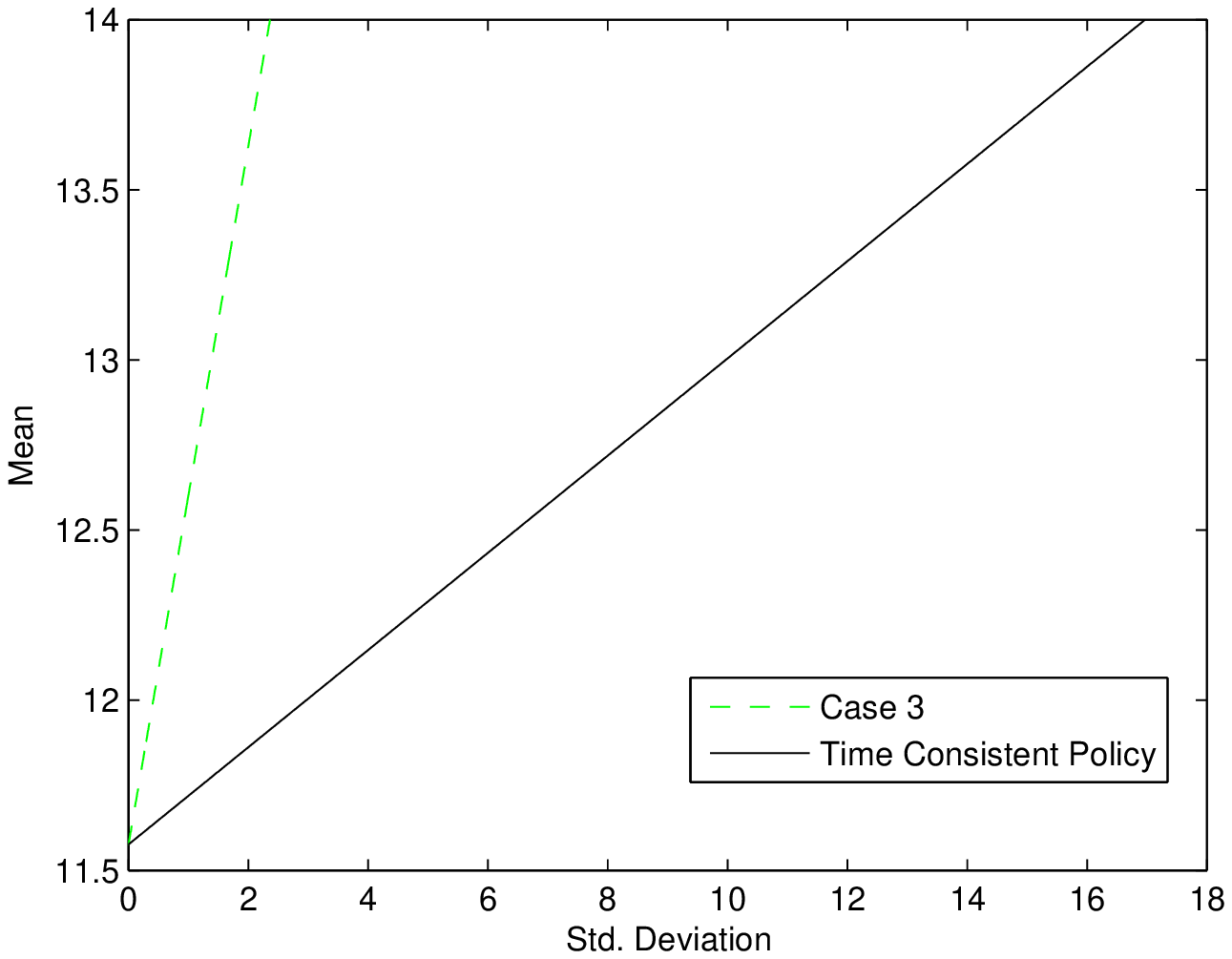}
   \end{minipage}}%
\caption{Comparison of efficient frontiers between Case 3 and the time consistent policy of Example \ref{ex3}} \label{fig:eff-frontiers-2}
\end{figure}

\end{exn}

\section{Conclusions}
We have developed in this paper a complete answer to the following question: Given a financial market with its return statistics known, what are the cone constraints on portfolio policies or what additional cone constraints are needed to be introduced such that the derived optimal portfolio policy is time consistent in efficiency. There are three main contributions of the paper: i) analytical solution of the mean-variance formulation for discrete-time cone constrained markets; ii) complete characterization of time consistency in efficiency and its close relationship to the minimum-variance signed supermartingale measure; and iii) a systematic framework in guaranteeing time consistency in efficiency by enforcing suitable cone constraints on portfolios.

More specifically, we have investigated in this paper the discrete-time mean-variance portfolio selection problem formulation in a convex cone constrained market, have given the condition under which there exists an admissible policy, have derived analytically the pre-committed efficient mean-variance policy, and have identified the explicit conditions under which the pre-committed efficient mean-variance policy is TCIE. The derived optimal policy is of a two-piece linear form, and this fact reveals that in a cone constrained market, mean-variance investors may switch between one efficient risky portfolio $\K_t^+$ and one inefficient risky portfolio $\K_t^-$ depending on the individual's current wealth level. Another prominent feature may also require our special attention: Market constraints make the current risky portfolios dependent not only on the current wealth level, but also on the future market conditions, even when the rates of return among different time periods are assumed to be independent.

Furthermore, we have extended the definition of variance-optimal signed martingale measure (VSMM) in unconstrained markets to minimum-variance signed supermartingale measure (VSSM) in constrained markets, and have derived the semi-analytical expression of VSSM's density (respect to the original probability measure), which only depends on the basic market setting (including the distribution of the excess rate of return, $\P_t$, and the set of portfolio constraints, $\mathcal{A}_t$). Our major finding demonstrates that the property of TCIE and VSSM are closely related, i.e., the pre-committed discrete-time efficient mean-variance policy (except for the minimum variance policy) satisfies TCIE if and only if the conditional expectation of VSSM's density is nonnegative, or once the conditional expectation becomes negative, it remains the same negative value until the terminal time. This interesting finding is the first analytical result that explicitly assesses the impact of constraints on the property of time consistency in dynamic decision problems and motivates us to establish a general solution framework in constructing TCIE dynamic portfolio selection models by introducing suitable portfolio constraints. The semi-analytical expression of VSSM's density may also benefit the research of mean-variance hedging in constrained markets.

An extension of our result to continuous-time cone constrained markets is straightforward, at least conceptually. On the other hand, if the rates of return among different periods are correlated, the problem will become more complicated and the idea of opportunity-neutral measure change in treating stochastic opportunity set in \cite{CernyKallsen2009} may be helpful. The real challenge appears when considering general markets with convex portfolio constraints (may not be a cone type). In such a market, the pre-committed efficient mean-variance policy may depend on more than two risky portfolios, making the analysis much more complicated.

\section*{Appendix:}

\subsection*{A1: The proof of Lemma \ref{lem:K-C}}

\proof From the definition in (\ref{eq:C-1}), it is easy to see
that $C_t^{\pm}>0$ for all $t=0,1,\dots,T-1$.

The first-order and second-order derivatives of $h_t^{\pm}(\K_t)$
with respect to $\K_t$ are given, respectively, as follows,
\begin{align*}
\bigtriangledown_{\K_t} h_t^{\pm}(\K_t) &= 2\E\left[C_{t+1}^+\Big(\P_t\P_t^{\prime}\K_t\mp\P_t\Big)1_{\{\P_t^{\prime}\K_t\leq \pm1\}}+C_{t+1}^-\Big(\P_t\P_t^{\prime}\K_t\mp\P_t\Big)1_{\{\P_t^{\prime}\K_t> \pm1\}}\right],\\
\bigtriangledown^2_{\K_t} h_t^{\pm}(\K_t) &= 2\E\left[C_{t+1}^+\P_t\P_t^{\prime}1_{\{\P_t^{\prime}\K_t\leq \pm1\}}+C_{t+1}^-\P_t\P_t^{\prime}1_{\{\P_t^{\prime}\K_t> \pm1\}}\right]\\
&\succeq 2\min(C_{t+1}^{+},C_{t+1}^{-})\E\left[\P_t\P_t^{\prime}\right]\\
&\succ 0.
\end{align*}
Therefore, $h_t^{\pm}(\K_t)$ are strictly convex with respect to
$\K_t$, which implies that $\K_t^{\pm}$ are uniquely determined.
Furthermore, $\K_t^{\pm}$ are optimal if and only if
\begin{align}\label{eqn:optimality-K}
\left(\bigtriangledown_{\K_t}
h_t^{\pm}(\K_t^{\pm})\right)^\prime(\K_t-\K_t^{\pm})\geq 0,
~~\forall~ \K_t\in \mathcal{A}_t,
\end{align}
(see Theorem 27.4 in \cite{Rockafellar}), which implies
\begin{align}
&\left(\bigtriangledown_{\K_t} h_t^{\pm}(\K_t^{\pm})\right)^\prime(\alpha\K_t^{\pm}-\K_t^{\pm})\geq 0,~~\forall~\alpha>0,\nonumber\\
&\left(\bigtriangledown_{\K_t}
h_t^{\pm}(\K_t^{\pm})\right)^\prime\K_t^{\pm}=
0,\label{eqn:optimality-K-reduce}
\end{align}
due to the assumption that $\mathcal{A}_t$ is a cone.

%

Then, we have
\begin{align*}
&\E\left[C_{t+1}^+\Big( 1\mp\P_t^{\prime}\K_t^{\pm}\Big)^21_{\{\P_t^{\prime}\K_t^{\pm}\leq \pm1\}}+C_{t+1}^-\Big( 1\mp\P_t^{\prime}\K_t^{\pm}\Big)^21_{\{\P_t^{\prime}\K_t^{\pm}> \pm1\}}\right]\\
=&\E\left[C_{t+1}^+\Big( 1\mp\P_t^{\prime}\K_t^{\pm}\Big)1_{\{\P_t^{\prime}\K_t^{\pm}\leq \pm1\}}+C_{t+1}^-\Big( 1\mp\P_t^{\prime}\K_t^{\pm}\Big)1_{\{\P_t^{\prime}\K_t^{\pm}> \pm1\}}\right]\\
&+\frac{1}{2}\left(\bigtriangledown_{\K_t} h_t^{\pm}(\K_t^{\pm})\right)^\prime\K_t^{\pm}\\
=&\E\left[C_{t+1}^+\Big(
1\mp\P_t^{\prime}\K_t^{\pm}\Big)1_{\{\P_t^{\prime}\K_t^{\pm}\leq
\pm1\}}+C_{t+1}^-\Big(
1\mp\P_t^{\prime}\K_t^{\pm}\Big)1_{\{\P_t^{\prime}\K_t^{\pm}>
\pm1\}}\right]
\end{align*}
and
\begin{align*}
&\E\left[C_{t+1}^+\Big( 1\mp\P_t^{\prime}\K_t^{\pm}\Big)^21_{\{\P_t^{\prime}\K_t^{\pm}\leq \pm1\}}+C_{t+1}^-\Big( 1\mp\P_t^{\prime}\K_t^{\pm}\Big)^21_{\{\P_t^{\prime}\K_t^{\pm}> \pm1\}}\right]\\
=&\E\left[C_{t+1}^+\Big( 1-(\K_t^{\pm})'\P_t\P_t^{\prime}\K_t^{\pm}\Big)1_{\{\P_t^{\prime}\K_t^{\pm}\leq \pm1\}}+C_{t+1}^-\Big( 1-(\K_t^{\pm})'\P_t\P_t^{\prime}\K_t^{\pm}\Big)1_{\{\P_t^{\prime}\K_t^{\pm}> \pm1\}}\right]\\
& +\left(\bigtriangledown_{\K_t} h_t^{\pm}(\K_t^{\pm})\right)^\prime\K_t^{\pm}\\
=&\E\left[C_{t+1}^+\Big(
1-(\K_t^{\pm})'\P_t\P_t^{\prime}\K_t^{\pm}\Big)1_{\{\P_t^{\prime}\K_t^{\pm}\leq
\pm1\}}+C_{t+1}^-\Big(
1-(\K_t^{\pm})'\P_t\P_t^{\prime}\K_t^{\pm}\Big)1_{\{\P_t^{\prime}\K_t^{\pm}>
\pm1\}}\right].
\end{align*}
Therefore,
\begin{align*}
C_t^+
=&\E\left[C_{t+1}^+\Big( 1-(\K_t^{+})'\P_t\P_t^{\prime}\K_t^{+}\Big)1_{\{\P_t^{\prime}\K_t^{+}\leq 1\}}+C_{t+1}^-\Big( 1-(\K_t^{+})'\P_t\P_t^{\prime}\K_t^{+}\Big)1_{\{\P_t^{\prime}\K_t^{+}> 1\}}\right]\\
\leq & \E\left[C_{t+1}^+\Big( 1-(\K_t^{+})'\P_t\P_t^{\prime}\K_t^{+}\Big)1_{\{\P_t^{\prime}\K_t^{+}\leq 1\}}\right]\\
\leq & C_{t+1}^{+}.
\end{align*}
The equality holds in the above inequality if and only if
$\K_t^+=\0$. The situation for $C_t^-$ can be proved similarly.
\endproof

\subsection*{A2: The proof of Theorem \ref{prop:cone-constained}}
\proof
Consider an auxiliary problem of $(P(d))$ by introducing
Lagrangian multiplier $2\mu$,
\begin{equation}\label{A_no_shorting}
\begin{array}{rl}
\min & \mathbb{E}\big[(x_T-d)^2+2\mu(x_T-d)\big], \\ [2mm]
\mbox{\rm s.t.}
& x_{t+1} =s_tx_t+\P_t'\u_t, \\ [2mm]
& \u_t\in\mathcal{A}_t, \quad t=0,1, \cdots, T-1,
\end{array}
\end{equation}
which is equivalent to the following formulation,
\[
\begin{array}{rl}
\min & \mathbb{E}\left[\displaystyle\frac{1}{2}\big(x_T-(d-\mu)\big)^2\right], \\ [2mm]
\mbox{\rm s.t.} & x_{t+1} =s_tx_t+\P_t'\u_t, \\ [2mm]
& \u_t\in\mathcal{A}_t, \quad t = 0, 1, \cdots, T-1.
\end{array}
\]
The above auxiliary problem can be further rewritten as
\begin{equation}\nonumber
\begin{array}{lll}
(L(\mu)): ~~~~ & \min & \mathbb{E}\left[\displaystyle\frac{1}{2}y_T^2\right], \\ [2mm]
& \mbox{\rm s.t.} & y_{t+1} =s_ty_t+\P_t'\u_t, \\ [2mm]
& & \u_t \in\mathcal{A}_t, \quad t = 0, 1, \cdots, T-1,
\end{array}
\end{equation}
where
\[
\begin{array}{l}
y_t \triangleq x_t-(d-\mu)\rho_t^{-1}, \quad t = 0, 1, \cdots, T.
\end{array}
\]

Now we will prove that the value function of $(L(\mu))$ at time $t$ is
\begin{align}
J_t(y_t)&=\min_{\mathbf{u}_t\in \mathcal{A}_t,\cdots,\mathbf{u}_{T-1}\in\mathcal{A}_{T-1}} \mathbb{E}\left[\frac{1}{2}y_T^2|\mathcal{F}_t\right]=\frac{1}{2}\rho_t^2\left[C_t^+y_t^21_{\{y_t\leq 0\}}+C_t^-y_t^21_{\{y_t> 0\}}\right], \label{opt-J}
\end{align}
where $C_t^+$ and $C_t^-$ are given in Lemma \ref{lem:K-C}.

At time $T$, we have
\begin{align*}
J_T(y_T)&=\frac{1}{2}y_T^2=\frac{1}{2}\rho_T^2\left[C_T^+y_T^21_{\{y_T\leq 0\}}+C_T^-y_T^21_{\{y_T> 0\}}\right].
\end{align*}
Thus, statement (\ref{opt-J}) holds true for time $T$. Assume
that statement (\ref{opt-J}) holds true for time $t+1$. We now
prove that the statement also remains true for time $t$. Applying the recursive relationship between $J_{t+1}$ and $J_t$ yields
\begin{equation}\label{eq:Jvalue}
\begin{array}{rl}
J_t(y_{t})
=\!\!\! & \displaystyle\min_{\u_t\in \mathcal{A}_t}\mathbb{E}[J_{t+1}(y_{t+1})|\mathcal{F}_t] \\ [2mm]
=\!\!\! & \displaystyle\min_{\u_t\in \mathcal{A}_t} \frac{1}{2}\rho_{t+1}^2 \mathbb{E}\Big[C_{t+1}^+y_{t+1}^21_{\{y_{t+1}\leq 0\}}+C_{t+1}^-y_{t+1}^21_{\{y_{t+1}> 0\}}|\mathcal{F}_t\Big]\\[2mm]
=\!\!\! & \displaystyle\min_{\u_t\in \mathcal{A}_t} \frac{1}{2}\rho_{t+1}^2 \mathbb{E}\Big[C_{t+1}^+(s_ty_t+\P_t'\u_t)^21_{\{\P_t'\u_t\leq -s_ty_t\}}+C_{t+1}^-(s_ty_t+\P_t'\u_t)^21_{\{\P_t'\u_t> -s_ty_t\}}|\mathcal{F}_t\Big].
\end{array}
\end{equation}

While $y_t<0$, identifying optimal $\u_t$ within the convex cone $\u_t\in\mathcal{A}_t$ is equivalent to identifying optimal $\K_t$ within the convex cone $\K_t\in\mathcal{A}_t$ when we set $\u_t=-s_t\K_ty_t$. We thus have
\[
J_t(y_t)=\min_{\K_t\in\mathcal{A}_t}\frac{1}{2}\rho_{t}^2 y_t^2 \E\left[C_{t+1}^+\Big( 1-\P_t^{\prime}\K_t\Big)^21_{\{\P_t^{\prime}\K_t\leq 1\}}+C_{t+1}^-\Big( 1-\P_t^{\prime}\K_t\Big)^21_{\{\P_t^{\prime}\K_t> 1\}}\right].
\]
From Lemma 1, the optimal control takes the following form,
\[
\u_t^{\star}=-s_t\K_t^{+}y_t.
\]
Substituting $\u^{\star}_t$ back to the value function (\ref{eq:Jvalue}) leads to
\begin{align*}
J_t(y_t)&=\displaystyle\frac{1}{2}\rho_{t}^2 y_t^2 \E\left[C_{t+1}^+\Big( 1-\P_t^{\prime}\K_t^+\Big)^21_{\{\P_t^{\prime}\K_t^+\leq 1\}}+C_{t+1}^-\Big( 1-\P_t^{\prime}\K_t^+\Big)^21_{\{\P_t^{\prime}\K_t^+> 1\}}\right]\\
&= \frac{1}{2}C_t^+\rho_t^2y_t^2.
\end{align*}

When $y_t>0$, identifying optimal $\u_t$ within the convex cone $\u_t\in\mathcal{A}_t$ is equivalent to identifying optimal $\K_t$ within the convex cone $\K_t\in\mathcal{A}_t$ when we set $\u_t=s_t\K_ty_t$. We thus have
\[
J_t(y_t)=\min_{\K_t\in\mathcal{A}_t}\frac{1}{2}\rho_{t}^2 y_t^2 \E\left[C_{t+1}^+\Big( 1+\P_t^{\prime}\K_t\Big)^21_{\{\P_t^{\prime}\K_t\leq -1\}}+C_{t+1}^-\Big( 1+\P_t^{\prime}\K_t\Big)^21_{\{\P_t^{\prime}\K_t> -1\}}\right].
\]
From Lemma 1, the optimal control takes the following form,
\[
\u_t^{\star}=s_t\K_t^{-}y_t.
\]
Substituting $\u^{\star}_t$ back to the value function (\ref{eq:Jvalue}) leads to
\begin{align*}
J_t(y_t)&=\displaystyle\frac{1}{2}\rho_{t}^2 y_t^2 \E\left[C_{t+1}^+\Big( 1+\P_t^{\prime}\K_t^-\Big)^21_{\{\P_t^{\prime}\K_t^-\leq -1\}}+C_{t+1}^-\Big( 1+\P_t^{\prime}\K_t^-\Big)^21_{\{\P_t^{\prime}\K_t^-> -1\}}\right]\\
&= \frac{1}{2}C_t^-\rho_t^2y_t^2.
\end{align*}

When $y_t=0$, we can
easily verify that $\u^\star_t = \0$ is the  minimizer. We can thus set
$$J_t(y_t)=\frac{1}{2}C_t^+\rho_{t}^2 y_t^2.$$

In summary, the optimal value for problem (\ref{A_no_shorting}) is
\begin{align}
\nonumber g(\mu)&=\min_{\u_0\in\mathcal{A}_0,\cdots,\u_{T-1}\in\mathcal{A}_{T-1}}\!\!\!\mathbb{E} \left[(x_T-d)^2+2\mu(x_T-d)\right]\\
&=\left\{\begin{array}{ll}
    C_0^+(d-\rho_0x_0-\mu)^2-\mu^2, & \mbox{if } \mu\leq d-\rho_0 x_{0}, \\ [2mm]
    C_0^-(d-\rho_0x_0-\mu)^2-\mu^2, & \mbox{if } \mu> d-\rho_0 x_{0},
  \end{array}\right.\label{eq:mu}
\end{align}
which is a first-order continuously differentiable concave function. To obtain the optimal value and optimal strategy for problem $(P(d))$, we maximize (\ref{eq:mu}) over $\mu\in\mathbb{R}$ according to Lagrangian duality theorem. We derive our results for three different value ranges of $d$.

i) $d=\rho_0x_0$.

The optimal Lagrangian multiplier takes zero value, i.e., $\mu^\star=0$. The optimal investment policy is thus $\u_t^\star=\0$, $t=0,1,\dots,T-1$.

ii) $d>\rho_0x_0$.

When $C_0^+=1$, i.e., $K_t^+=\0$, $t=0,1,\dots,T-1$, we can take $\mu^\star=-\infty$ resulting $g(\mu^\star)=+\infty$. This means that $P(d)$ does not have a feasible solution. When $C_0^+<1$ and $C_0^-=1$, $C_0^+(d-\rho_0x_0-\mu)^2-\mu^2$ is a strictly concave function and $C_0^-(d-\rho_0x_0-\mu)^2-\mu^2$ is a decreasing linear function. The optimal Lagrangian multiplier satisfies
\begin{align*}
\mu^\star=\frac{d-\rho_0x_0}{1-(C_0^+)^{-1}}<(d-\rho_0x_0).
\end{align*}
When $C_0^+<1$ and $C_0^-<1$, $C_0^{\pm}(d-\rho_0x_0-\mu)^2-\mu^2$ are both strictly concave.
The optimal Lagrangian multiplier satisfies
\begin{align*}
\mu^\star=\frac{d-\rho_0x_0}{1-(C_0^+)^{-1}}<(d-\rho_0x_0).
\end{align*}
Therefore, the optimal mean-variance pair is presented by
\[
\big(\E[x_T], \ \mbox{\rm Var}(x_T)\big) = (d,~g(\mu^\star)) =\left(d,~ \frac{C_{0}^{+}\left(d-\rho_0 x_{0}\right)^2}{1-C_{0}^{+}}\right).
\]

iii) $d<\rho_0x_0$.

Similarly, when $C_0^-=1$, $P(d)$ does not have a feasible solution. When $C_0^-<1$, the optimal Lagrangian multiplier satisfies
\begin{align*}
\mu^\star=\frac{d-\rho_0x_0}{1-(C_0^-)^{-1}}>(d-\rho_0x_0).
\end{align*}
Then, the optimal mean-variance pair is presented by
\[
\big(\E[x_T], \ \mbox{\rm Var}(x_T)\big) = (d,~g(\mu^\star)) =\left(d,~ \frac{C_{0}^{-}\left(d-\rho_0 x_{0}\right)^2}{1-C_{0}^{-}}\right).
\]
Therefore, $g(\mu)$ attains its maximum value at $\mu^\star$ expressed in (\ref{eq:opt-mu}).
Moreover, the optimal mean-variance pair of problem $(P(d))$ is presented by
\[
\big(\E[x_T], \ \mbox{\rm Var}(x_T)\big) = \left(d,~ \frac{C_{0}^{+}\left(d-\rho_0 x_{0}\right)^2}{1-C_{0}^{+}}1_{\{d\geq \rho_0 x_0\}}+\frac{C_{0}^{-}\left(d-\rho_0 x_{0}\right)^2}{1-C_{0}^{-}}1_{\{d< \rho_0 x_0\}}\right).
\]
Finally, the efficient frontier follows naturally from our above discussion.

\endproof

\subsection*{A3: The proof of Theorem \ref{prop:truncated-efficency}}

\proof Comparing Eq. (\ref{eq:opt-u-star}) with Eq.
(\ref{eq:opt-u-star-truncated}), we can conclude that at time $k$,
the truncated pre-committed efficient mean-variance policy,
$\u_t^{\star}, t=k,k+1,\dots,T-1$, also solves $(P(d_k) \mid x_k)$ when
$d_k$ satisfies $d-\mu^{\star}=d_k-\mu_k^{\star}$. Note from the
discussion after Theorem \ref{prop:cone-constained} that the
solution to $(P(d_k) \mid x_k)$ is inefficient if and only if $d_k < \rho_k
x_{k}$ and $C_k^-<1$ (or equivalently, the solution to $(P(d_k) \mid x_k)$
is efficient if i) $d_k\geq \rho_k x_{k}$, or ii) $d_k <\rho_k
x_{k}$ and $C_k^-=1$). When $0<C_k^+<1$, we have
\begin{align*}
& ~d_k\geq \rho_k x_{k}\\
\Leftrightarrow & ~(d_k-\rho_k x_{k})\frac{1}{1-C_k^+}\geq 0\\
\Leftrightarrow & ~d_k-\frac{d_k-\rho_k x_{k}}{1-(C_k^+)^{-1}}\geq \rho_k x_k\\
\Leftrightarrow&~d_k-\mu_k^\star\geq \rho_k x_k, ~\mbox{if } d_k\geq \rho_k x_{k}\\
\Leftrightarrow&~d_k-\mu_k^\star\geq \rho_k x_k,\\
\Leftrightarrow&~d-\mu^\star\geq \rho_k x_k.
\end{align*}
Therefore, when both $d-\mu^\star\geq \rho_k x_k$ and $C_k^+<1$
hold, the truncated pre-committed efficient mean-variance policy
remains efficient for the truncated problem $(P(d_k) \mid x_k)$.


Similarly, when $0<C_k^-<1$, we have
\begin{align*}
& ~d_k < \rho_k x_{k}\\
\Leftrightarrow & ~(d_k-\rho_k x_{k})\frac{1}{1-C_k^-} < 0\\
\Leftrightarrow & ~d_k-\frac{d_k-\rho_k x_{k}}{1-(C_k^-)^{-1}} < \rho_k x_k\\
\Leftrightarrow&~d_k-\mu_k^\star < \rho_k x_k, ~\mbox{if } d_k < \rho_k x_{k}\\
\Leftrightarrow&~d_k-\mu_k^\star < \rho_k x_k,\\
\Leftrightarrow&~d-\mu^\star < \rho_k x_k,
\end{align*}
which implies that when both $d-\mu^\star< \rho_k x_k$ and
$C_k^-<1$ hold, the truncated pre-committed efficient
mean-variance policy switches to be inefficient for the truncated
problem $(P(d_k) \mid x_k)$.

When $d-\mu^\star\geq \rho_k x_k$, $C_k^+=1$ or $d-\mu^\star<
\rho_k x_k$, $C_k^-=1$ hold, we have $\u_t^\star=\0$,
$t=k,k+1,\dots,T-1$, i.e., the truncated pre-committed efficient
mean-variance policy becomes the minimum variance policy for the
truncated problem $(P(d_k) \mid x_k)$.

The proposition follows when combining the results for all the
situations discussed above.
\endproof

\subsection*{A4: The proof of Proposition \ref{prop:probability-issue}}

\proof We only need to prove the first, the third and the fifth
equalities.

 Condition$(d-\mu^\star)>\rho_{t}x_{t}$ dictates the optimal policy at time $t$ as $\u_t^{\star}=s_t\K_t^+\big((d_k-\mu^{\star}_k)\rho_t^{-1}-x_t\big)$. The wealth level at time $t+1$ follows
\begin{align*}
x_{t+1}=s_tx_t+s_t\P_t^\prime\K_t^+\big((d_k-\mu^{\star}_k)\rho_t^{-1}-x_t\big),
\end{align*}
which implies
\begin{align*}
&~(d-\mu^\star)\geq\rho_{t+1}x_{t+1}\\
\Leftrightarrow&~(d-\mu^\star)\geq\rho_{t}x_{t}+\rho_t\P_t^\prime\K_t^+\big((d_k-\mu^{\star}_k)\rho_t^{-1}-x_t\big)\\
\Leftrightarrow&~\left[(d-\mu^\star)-\rho_{t}x_{t}\right](1-\P_t^\prime\K_t^+)\geq 0\\
\Leftrightarrow&~\P_t^\prime\K_t^+\leq 1.
\end{align*}
Thus the first statement holds.

 Condition $(d-\mu^\star)<\rho_{t}x_{t}$ dictates the optimal policy at time $t$ as $\u_t^{\star}=-s_t\K_t^-\big((d_k-\mu^{\star}_k)\rho_t^{-1}-x_t\big)$. The wealth level at time $t+1$ is
\begin{align*}
x_{t+1}=s_tx_t-s_t\P_t^\prime\K_t^-\big((d_k-\mu^{\star}_k)\rho_t^{-1}-x_t\big),
\end{align*}
which implies
\begin{align*}
&~(d-\mu^\star)\geq\rho_{t+1}x_{t+1}\\
\Leftrightarrow&~(d-\mu^\star)\geq\rho_{t}x_{t}-\rho_t\P_t^\prime\K_t^-\big((d_k-\mu^{\star}_k)\rho_t^{-1}-x_t\big)\\
\Leftrightarrow&~\left[(d-\mu^\star)-\rho_{t}x_{t}\right](1+\P_t^\prime\K_t^-)\geq 0\\
\Leftrightarrow&~\P_t^\prime\K_t^-\leq -1.
\end{align*}
Thus the third statement holds.

 Condition $(d-\mu^\star)=\rho_{t}x_{t}$ dictates the optimal policy at time $t$ as $\u_t^{\star}=\0$. The wealth level at time $t+1$ is
\begin{align*}
x_{t+1}=s_tx_t,
\end{align*}
which implies
\begin{align*}
&~(d-\mu^\star)=\rho_{t+1}x_{t+1}=\rho_{t}x_{t}.
\end{align*}
Thus the fifth statement holds.
\endproof

\subsection*{A5: The proof of Lemma \ref{lem:vssm}}

\proof We solve both problems by duality theory. The dual problem
of $(A^+(t))$ is
\begin{equation*}
\max_{\nu_{t}\in\mathbb{R}}\max_{-\lambda_{t}\in\mathcal{A}_{t}}\mathbb{E}\left[\min_{m_{t+1}}L_t(m_{t+1},\nu_t,\lambda_t)\Big|\mathcal{F}_{t}
\right],
\end{equation*}
where the Lagrangian function is defined as
\begin{align*}
L_t(m_{t+1},\nu_t,\lambda_t)\triangleq\left(\frac{1}{C_{t+1}^+}1_{\{m_{t+1}\geq
0\}}+\frac{1}{C_{t+1}^-}1_{\{m_{t+1}< 0\}}\right)
m_{t+1}^2-\nu_{t} m_{t+1}+\nu_{t}-\lambda_{t}^\prime\P_{t}m_{t+1}.
\end{align*}
We also define
\begin{align*}
\mathcal{D}(\nu_t,\lambda_t)\triangleq\mathbb{E}\left[\min_{m_{t+1}}L_t(m_{t+1},\nu_t,\lambda_t)\Big|\mathcal{F}_{t}
\right].
\end{align*}
The first order condition of $L_t(m_{t+1},\nu_t,\lambda_t)$ with
respect to $m_{t+1}$ gives rise to
\begin{align}
m_{t+1}=\frac{C_{t+1}^+}{2}(\nu_t+\lambda_t^\prime\P_t)1_{\{\nu_t+\lambda_t^\prime\P_t\geq
0\}}+\frac{C_{t+1}^-}{2}(\nu_t+\lambda_t^\prime\P_t)1_{\{\nu_t+\lambda_t^\prime\P_t<
0\}}.\label{eqn:m}
\end{align}
Note that $m_{t+1}\geq 0$ if and only if
$\nu_t+\lambda_t^\prime\P_t\geq 0$.

Then we have
\begin{align*}
\mathcal{D}(\nu_t,\lambda_t)=\mathbb{E}\left[-\frac{1}{4}(\nu_t+\lambda_t^\prime\P_t)^2\left(C_{t+1}^+1_{\{\nu_t+\lambda_t^\prime\P_t\geq
0\}}+C_{t+1}^-1_{\{\nu_t+\lambda_t^\prime\P_t<
0\}}\right)+\nu_t\right].
\end{align*}
If $\nu_t>0$, identifying optimal $\lambda_{t}$ within the convex
cone $-\lambda_{t}\in \mathcal{A}_{t}$ is equivalent to
identifying optimal $\K_t$ within the convex cone
$\K_t\in\mathcal{A}_t$ when we set $\lambda_t=- \nu_t \K_t$. Then,
\begin{align*}
&\max_{\nu_{t}>0}\max_{-\lambda_{t}\in\mathcal{A}_{t}}\mathcal{D}(\lambda_t,\nu_t) \\
=& \max_{\nu_{t}>0}\max_{\K_{t}\in\mathcal{A}_{t}}\mathbb{E}\left[-\frac{1}{4}\nu_t^2(1-\K_t^\prime\P_t)^2\left(C_{t+1}^+1_{\{\K_t^\prime\P_t\leq 1\}}+C_{t+1}^-1_{\{\K_t^\prime\P_t>1 \}}\right)+\nu_t\right]\\
=&\max_{\nu_{t}>0} \left\{-\frac{1}{4}\nu_t^2
\left\{\min_{\K_{t}\in\mathcal{A}_{t}}
\mathbb{E}\left[(1-\K_t^\prime\P_t)^2\left(C_{t+1}^+1_{\{\K_t^\prime\P_t\leq
1\}}+C_{t+1}^-1_{\{\K_t^\prime\P_t>1
\}}\right)\right]\right\}+\nu_t\right\}.
\end{align*}
Therefore, $\mathcal{D}(\lambda_t,\nu_t)$ attains its maximum
$\frac{1}{C_t^+}$ at
\begin{align}
\lambda_t^+&=-\nu_t\K_t^+,\label{eqn:m-lamda} \\
\nu_t^+&=\frac{2}{C_t^+}.\label{eqn:m-mu}
\end{align}
If $\nu_t<0$, identifying optimal $\lambda_{t}$ within the convex
cone $-\lambda_{t}\in\mathcal{A}_{t}$ is equivalent to identifying
optimal $\K_t$ within the convex cone $\K_t\in\mathcal{A}_t$ when
we set $\lambda_t=\nu_t \K_t$. Then,
\begin{align*}
&\max_{\nu_{t}<0}\max_{-\lambda_{t}\in\mathcal{A}_{t}}\mathcal{D}(\lambda_t,\nu_t) \\
=& \max_{\nu_{t}<0}\max_{\K_{t}\in\mathcal{A}_{t}}\mathbb{E}\left[-\frac{1}{4}\nu_t^2(1-\K_t^\prime\P_t)^2\left(C_{t+1}^+1_{\{\K_t^\prime\P_t\leq -1\}}+C_{t+1}^-1_{\{\K_t^\prime\P_t>-1 \}}\right)+\nu_t\right]\\
=&\max_{\nu_{t}<0} \left\{-\frac{1}{4}\nu_t^2
\left\{\min_{\K_{t}\in\mathcal{A}_{t}}
\mathbb{E}\left[(1-\K_t^\prime\P_t)^2\left(C_{t+1}^+1_{\{\K_t^\prime\P_t\leq
-1\}}+C_{t+1}^-1_{\{\K_t^\prime\P_t>-1
\}}\right)\right]\right\}+\nu_t\right\}.
\end{align*}
Now, $\mathcal{D}(\lambda_t,\nu_t)$ attains its maximum $0$ when
$\nu_t\uparrow 0$.

Substituting both (\ref{eqn:m-lamda}) and (\ref{eqn:m-mu}) into
(\ref{eqn:m}) yields the expression of $m_{t+1}^+$,
\begin{align*}
m_{t+1}^+&=\frac{1}{C_{t}^+}\left[C_{t+1}^+(1-\P_t^\prime\K_t^{+})1_{\{\P_t^\prime\K_t^{+}\leq 1\}}+C_{t+1}^-(1-\P_t^\prime\K_t^{+})1_{\{\P_t^\prime\K_t^{+}> 1\}}\right],\\
&=\frac{1}{C_{t}^+}\left[C_{t+1}^+(1-\P_t^\prime\K_t^{+})1_{\{m_{t+1}^+\geq
0\}}+C_{t+1}^-(1-\P_t^\prime\K_t^{+})1_{\{m_{t+1}^+< 0\}}\right],
\end{align*}
and the optimal objective value of $(A^+(t))$,
\begin{align*}
&\mathbb{E}\left[\left(\displaystyle\frac{1}{C_{t+1}^+}1_{\{m_{t+1}^+\geq 0\}}+\frac{1}{C_{t+1}^-}1_{\{m_{t+1}^+< 0\}}\right) (m_{t+1}^+)^2\Big|\mathcal{F}_{t}\right]\\
=&\frac{1}{(C_{t}^+)^2}\mathbb{E}\left[C_{t+1}^+(1-\P_t^\prime\K_t^{+})^21_{\{\P_t^\prime\K_t^{+}\leq 1\}}+C_{t+1}^-(1-\P_t^\prime\K_t^{+})^21_{\{\P_t^\prime\K_t^{+}>1\}}\right]\\
=&\frac{1}{C_{t}^+}.
\end{align*}
Notice that $m_{t+1}^+\geq 0$ if and only if
$\P_t^\prime\K_t^{+}\leq 1$.

Applying a similar approach to problem $(A^-(t))$ gives rise to
the expression of $m_{t+1}^-$ and the corresponding optimal
optimal value $\frac{1}{C_{t}^-}$. Notice that $m_{t+1}^-\leq 0$
if and only if $\P_t^\prime\K_t^{-}\leq -1$.
\endproof

\subsection*{A6: The proof of Theorem \ref{prop:VSSM} }

\proof The problem of finding the density of the minimum variance
signed supermartingale measure is formulated as
\begin{equation}\label{Prob_VSSM}
\begin{array}{rrl}
(P_{VSSM}):~~~~~ & \min & \mathbb{E}\left[m_1^2m_2^2\cdots m_{T}^2
\right] \\ [2mm] & \mbox{\rm s.t.} &
\mathbb{E}\left[m_{t+1}\big|\mathcal{F}_t\right] = 1, \\ [2mm] & &
\mathbb{E}\big[m_1m_2\cdots
m_{T}\P_{t}\big|\mathcal{F}_t\big]\in\mathcal{A}_t^\perp, \\ [2mm]
& & m_{t+1}\in\mathcal{L}^2(\mathcal{F}_{t+1},P),\\[2mm]
& & t = 0, 1, \cdots, T-1.
\end{array}
\end{equation}
We will prove by induction that the cost-to-go function of
$(P_{VSSM})$ at time $t$ is given by
\begin{align*}
J(m_1m_2\dots m_{t})=\frac{1}{C_t^+}m_1^2m_2^2\cdots
m_{t}^21_{\{m_1m_2\cdots m_{t}\geq
0\}}+\frac{1}{C_t^-}m_1^2m_2^2\cdots m_{t}^21_{\{m_1m_2\cdots
m_{t}< 0\}},
\end{align*}
which implies (\ref{eq:second-monent-VSSM}).

At time $T$, the statement holds true by recognizing $C_T^\pm=1$.
Assume that the statement holds true for time $t+1$. We now prove
that the statement also remains true for time $t$.

At time $t$, when $m_1m_2\cdots m_{t}> 0$, $(P_{VSSM})$ reduces to
\begin{equation*}
\begin{array}{rrl}
& \min & m_1^2m_2^2\cdots
m_{t}^2\mathbb{E}\left[\left(\displaystyle\frac{1}{C_{t+1}^+}1_{\{m_{t+1}\geq
0\}}+\frac{1}{C_{t+1}^-}1_{\{m_{t+1}<
0\}}\right)m_{t+1}^2\Big|\mathcal{F}_{t} \right] \\ [2mm] &
\mbox{\rm s.t.} & \mathbb{E}[m_{t+1}|\mathcal{F}_{t} ] = 1, \\
[2mm]
& & \mathbb{E}\big[m_{t+1}\P_{t}\big|\mathcal{F}_{t}\big]\in\mathcal{A}_{t}^\perp,\\[2mm]
& & m_{t+1}\in\mathcal{L}^2(\mathcal{F}_{t+1},P).
\end{array}
\end{equation*}
On the other hand, when $m_1m_2\cdots m_{t}< 0$, $(P_{VSSM})$
reduces to
\begin{equation*}
\begin{array}{rrl}
& \min & m_1^2m_2^2\cdots
m_{t}^2\mathbb{E}\left[\left(\displaystyle\frac{1}{C_{t+1}^+}1_{\{m_{t+1}\leq
0\}}+\frac{1}{C_{t+1}^-}1_{\{m_{t+1}>
0\}}\right)m_{t+1}^2\Big|\mathcal{F}_{t} \right] \\ [2mm] &
\mbox{\rm s.t.} & \mathbb{E}[m_{t+1}|\mathcal{F}_{t} ] = 1, \\
[2mm]
& & -\mathbb{E}\big[m_{t+1}\P_{t}\big|\mathcal{F}_{t}\big]\in\mathcal{A}_{t}^\perp,\\[2mm]
& & m_{t+1}\in\mathcal{L}^2(\mathcal{F}_{t+1},P).
\end{array}
\end{equation*}

With the help of Lemma \ref{lem:vssm}, the optimal solution is
\begin{align*}
m_{t+1}^*&=m_{t+1}^+1_{\{m_1m_2\cdots m_{t}\geq
0\}}+m_{t+1}^-1_{\{m_1m_2\cdots m_{t}< 0\}}.
\end{align*}
(When $m_1m_2\cdots m_{t}= 0$, we can set $m_{t+1}^*=m_{t+1}^+$.)

Then, the cost-to-go function becomes
\begin{align*}
J(m_1m_2\dots m_{t})=\frac{1}{C_t^+}m_1^2m_2^2\cdots
m_{t}^21_{\{m_1m_2\cdots m_{t}\geq
0\}}+\frac{1}{C_t^-}m_1^2m_2^2\cdots m_{t}^21_{\{m_1m_2\cdots
m_{t}< 0\}}.
\end{align*}

Now the remaining part in our proof is to prove that
\begin{align*}
m_1^*m_2^*\cdots m_{T}^*=(C_0^+)^{-1}\prod_{i=0}^{T-1}B_i.
\end{align*}
We will prove that
\begin{align*}
m_1^*m_2^*\cdots
m_{t}^*=(C_0^+)^{-1}\prod_{i=0}^{t-1}B_i\left(C_t^+1_{\{m_1^*m_2^*\cdots
m_{t}^*\geq 0\}}+C_t^-1_{\{m_1^*m_2^*\cdots m_{t}^*< 0\}}\right),
\end{align*}
which implies the conditional expectation in
(\ref{eq:conditional-expectation-VSSM}).

When $t=1$, the following is obvious,
\begin{align*}
m_1^*=m_1^+=(C_0^+)^{-1}C_1^+(1-\P_0^\prime\K_0^+)=(C_0^+)^{-1}\prod_{i=0}^{0}B_i\left(C_1^+\right).
\end{align*}
Assume that at time $t$ our statement holds true, we prove now
that the statement also holds for time $t+1$, as
\begin{align*}
&m_1^*m_2^*\cdots m_{t+1}^*\\
=&(C_0^+)^{-1}\prod_{i=0}^{t-1}B_i\left(C_t^+1_{\{m_1^*\cdots m_{t}^*\geq 0\}}+C_t^-1_{\{m_1^*\cdots m_{t}^*< 0\}}\right)\left(m_{t+1}^+1_{\{m_1^*\cdots m_{t}^*\geq 0\}}+m_{t+1}^-1_{\{m_1^*\cdots m_{t}^*< 0\}}\right)\\
=&(C_0^+)^{-1}\prod_{i=0}^{t-1}B_i\big[C_{t+1}^+(1-\P_t^\prime\K_t^+)1_{\{m_1^*\cdots m_t^*\geq 0\}}1_{\{m_{t+1}^+\geq 0\}}+C_{t+1}^-(1-\P_t^\prime\K_t^+)1_{\{m_1^*\cdots m_t^*\geq 0\}}1_{\{m_{t+1}^+< 0\}}\\
&+C_{t+1}^+(1+\P_t^\prime\K_t^-)1_{\{m_1^*\cdots m_t^*< 0\}}1_{\{m_{t+1}^-\leq 0\}}+C_{t+1}^-(1+\P_t^\prime\K_t^-)1_{\{m_1^*\cdots m_t^*< 0\}}1_{\{m_{t+1}^-> 0\}}\big]\\
=&(C_0^+)^{-1}\prod_{i=0}^{t-1}B_i\left[(1-\P_t^\prime\K_t^+)1_{\{m_1^*\cdots m_t^*\geq 0\}}+(1+\P_t^\prime\K_t^-)1_{\{m_1^*\cdots m_t^*< 0\}}\right]\\
&\cdot\left(C_{t+1}^+1_{\{m_1^*\cdots m_{t+1}^*\geq 0\}}+C_{t+1}^-1_{\{m_1^*\cdots m_{t+1}^*< 0\}}\right)\\
=&(C_0^+)^{-1}\prod_{i=0}^{t}B_i\left(C_{t+1}^+1_{\{m_1^*\cdots
m_{t+1}^*\geq 0\}}+C_{t+1}^-1_{\{m_1^*\cdots m_{t+1}^*<
0\}}\right).
\end{align*}
Therefore
\begin{align*}
m_1^*m_2^*\cdots
m_{T}^*=(C_0^+)^{-1}\prod_{i=0}^{T-1}B_i\left(C_T^+1_{\{m_1^*m_2^*\cdots
m_T^*\geq 0\}}+C_T^-1_{\{m_1^*m_2^*\cdots m_T^*<
0\}}\right)=(C_0^+)^{-1}\prod_{i=0}^{T-1}B_i.
\end{align*}
\endproof

\subsection*{A7: The proof of Theorem \ref{thm:TCIE-VSSM}}

\proof Lemma \ref{lem:TCIE-condition} already states the necessary
and sufficient condition under which the pre-committed efficient
policy (except for the minimum variance policy) satisfies TCIE,
which can be summarized as follows:

For $t=1,2,\cdots,T-1$,
\begin{align*}
&\forall ~ x_t^\star,~~ d-\mu^{\star}\geq \rho_tx_t^\star;\\
\mbox{ or } & \exists ~ x_t^\star, ~~ d-\mu^{\star}<
\rho_tx_t^\star,~C_t^-=1;
\end{align*}
with $d-\mu^{\star}> \rho_0x_0$.

If at time $t$, $\exists ~ x_t^\star, ~ d-\mu^{\star}<
\rho_tx_t^\star,~C_t^-=1$, then
\begin{align*}
 ~ x_{t+1}^\star=s_tx_t^\star, ~~ d-\mu^{\star}< \rho_tx_t^\star=\rho_{t+1}x_{t+1}^\star,~C_{t+1}^-=1,
\end{align*}
if and only if $\K_i^-=\0$, ($\u_i^\star=\0$),
$i=t,t+1,\cdots,T-1$.

Therefore, the necessary and sufficient condition can be
reexpressed as
\begin{align*}
&\forall~ T>t>0,~\forall ~ x_t^\star,~~d-\mu^{\star}\geq \rho_tx_t^\star;\\
\mbox{ or } & C_{\tau}^-=1, \mbox{ where }\tau=\inf\left\{~t~
\big\vert ~d-\mu^{\star}< \rho_tx_t^\star, ~~t = 1, 2, \cdots,
T\right\}
\end{align*}
with $d-\mu^{\star}> \rho_0x_0$.

Based on the expression of $x_t^\star$ in
(\ref{eq:x-star-discrete}), we have
\begin{align*}
d-\mu^{\star}\geq (<) ~\rho_tx_t^\star ~~\Leftrightarrow ~~[(d -
\mu^\star) - x_0\rho_0]\prod_{i=0}^{t-1}B_i\geq (<)
~0~~\Leftrightarrow ~~\prod_{i=0}^{t-1}B_i\geq (<) ~0.
\end{align*}
Furthermore, for $T\geq k\geq \tau$ and $T>j\geq\tau$,
\begin{align*}
&~C_{\tau}^-=1,~~\prod_{i=0}^{\tau-1}B_i< 0,\\
\Leftrightarrow&~\K_j^-=\0,~~C_k^-=1,~~\prod_{i=0}^{\tau-1}B_i< 0,\\
\Leftrightarrow&\prod_{i=0}^{k-1}B_i\left(C_k^+1_{\{\prod_{i=0}^{k-1}B_i\geq
0\}}+C_k^-1_{\{\prod_{i=0}^{k-1}B_i< 0\}}\right)
=\prod_{i=0}^{\tau-1}B_i\left(C_\tau^+1_{\{\prod_{i=0}^{\tau-1}B_i\geq
0\}}+C_\tau^-1_{\{\prod_{i=0}^{\tau-1}B_i< 0\}}\right)<0.
\end{align*}
Noticing the expression of
$\mathbb{E}\left[\frac{d\tilde{P}}{dP}\Big|\mathcal{F}_t\right]$
in (\ref{eq:conditional-expectation-VSSM}) finally concludes our
proof.
\endproof

\subsection*{A8: The proof of Theorem \ref{prop:condition-TCIE}}

\proof Under the condition in the proposition, we have
\begin{align*}
\left(\bigtriangledown_{\K_t} h_t^{-}(\0)\right)^\prime(\K_t-\0)=2C_{t+1}^-\E[\P_t^\prime]\K_t\geq 0,~\forall~ \K_t\in\mathcal{A}_t,
\end{align*}
which implies $\K_t^-=\0$ and $C_t^-=1$ for all $t$.
\endproof

\subsection*{A9: The time consistent policy of $(P(d))$}
In the solution framework proposed by \cite{BasakChabakauri} and \cite{Bjork:2012}, the so-called time consistent policy at time $t$ is derived by a backward induction, taking into account that optimal investment decisions have already been taken in the future. Thus, the time consistent policy is the collection of equilibrium strategies adopted by fictitious investors at different times in a sequential game. More specifically, the time $t$ investor considers the following problem,
\begin{equation}\label{prob:cone_constrained_TC}
\begin{array}{rl}
(P_t(d)): & \quad \left\{\begin{array}{rl}
\displaystyle\min_{\u_t\in\mathbb{R}^n} & \mbox{Var}(x_T|x_t)\equiv \mathbb{E}\big[(x_T-d)^2 |x_t\big], \\ [2mm]
\mbox{\rm s.t.} & \mathbb{E}[x_T|x_t] = d, \\ [2mm]
& x_{t+1} =s_tx_t+\mathbf{P}_t'\mathbf{u}_t, \\ [2mm]
& x_{j+1} =s_jx_j+\mathbf{P}_j'\bar{\mathbf{u}}_j, \quad j = t+1, \cdots, T-1,\\ [2mm]
& \bar \u_j~~\mbox{solves Problem } (P_j(d)), \quad j = t+1, \cdots, T-1.
\end{array}\right.
\end{array}
\end{equation}
We will prove by induction that the time consistent policy, the conditional mean and conditional variance of terminal wealth under time consistent policy are given as
\begin{align}
\label{eq:tc-u}&\bar\u_t=-\E^{-1}[\P_t\P_t']\E[\P_t]\frac{x_t\rho_t-d}{B_t\rho_{t+1}},\\
\label{eq:tc-mean}&\E[x_T|x_t]_{\{\bar\u\}}=d,\\
\label{eq:tc-variance}&\mbox{Var}(x_T|x_t)_{\{\bar\u\}}=(d-x_t\rho_t)^2D_t,
\end{align}
where $B_t=\E[\P_t']\E^{-1}[\P_t\P_t']\E[\P_t]$ and
\begin{align*}
D_t=\prod_{j=t}^{T-1}\frac{1-B_j}{B_j}> 0.
\end{align*}

We start our proof from time $T-1$ where the investor faces the following optimization problem,
\begin{equation*}
\begin{array}{rl}
(P_{T-1}(d)): & \quad \left\{\begin{array}{cl}
\displaystyle\min_{\u_{T-1}\in\mathbb{R}^n} & \mathbb{E}\big[(s_{T-1}x_{T-1}+\mathbf{P}_{T-1}'\mathbf{u}_{T-1}-d)^2 |x_{T-1}\big], \\ [2mm]
\mbox{\rm s.t.} & \mathbb{E}[s_{T-1}x_{T-1}+\mathbf{P}_{T-1}'\mathbf{u}_{T-1}|x_{T-1}] = d,
\end{array}\right.
\end{array}
\end{equation*}
which can be solved by the Lagrangian method with its solution given as
\begin{align*}
&\bar\u_{T-1}=-\E^{-1}[\P_{T-1}\P_{T-1}^{\prime}]\E[\P_{T-1}]\frac{x_{T-1}\rho_{T-1}-d}{B_{T-1}},\\
&\E[x_T|x_{T-1}]_{\{\bar\u\}}=d,\\
&\mbox{Var}(x_T|x_{T-1})_{\{\bar\u\}}=(d-x_{T-1}\rho_{T-1})^2D_{T-1}.
\end{align*}
Assume that at time $t+1$, (\ref{eq:tc-u}), (\ref{eq:tc-mean}) and (\ref{eq:tc-variance}) hold. Then at time $t$, the investor faces the following optimization problem,
\begin{equation*}
\begin{array}{rl}
(P_{t}(d)): & \quad \left\{\begin{array}{cl}
\displaystyle\min_{\u_{t}\in\mathbb{R}^n} & \mbox{Var}(x_T|x_t)\equiv\E[\mbox{Var}(x_T|x_{t+1})_{\{\bar\u\}} |x_{t}\big]+\mbox{Var}\big(\E[x_T|x_{t+1}]_{\{\bar\u\}}|x_{t}\big), \\ [2mm]
\mbox{\rm s.t.} & \mathbb{E}[\E[x_T|x_{t+1}]_{\{\bar\u\}}|x_{t}] = d,\\[2mm]
& x_{t+1} =s_tx_t+\mathbf{P}_t'\mathbf{u}_t,
\end{array}\right.
\end{array}
\end{equation*}
which is equivalent to
\begin{equation*}
\displaystyle\min_{\u_{t}\in\mathbb{R}^n}\quad D_{t+1}\E[(d-(s_tx_t+\mathbf{P}_t'\mathbf{u}_t)\rho_{t+1})^2 |x_{t}\big].
\end{equation*}
It is not difficult to verify the following optimal solutions for $(P_{t}(d))$,
\begin{align*}
&\bar\u_{t}=-\E^{-1}[\P_{t}\P_{t}^{\prime}]\E[\P_{t}]\frac{x_{t}\rho_{t}-d}{B_{t}\rho_{t+1}},\\
&\E[x_T|x_{t}]_{\{\bar\u\}}=d,\\
&\mbox{Var}(x_T|x_{t})_{\{\bar\u\}}=(d-x_{t}\rho_{t})^2D_{t}.
\end{align*}
Therefore, the efficient frontier of the time consistent policy is given as
\begin{align}\label{eq:eff-frontier-TC}
\mbox{Var}(x_T)_{\{\bar\u\}}=(\E[x_T]_{\{\bar\u\}}-x_{0}\rho_{0})^2D_{0},\quad \E[x_T]_{\{\bar\u\}}\geq x_{0}\rho_{0}.
\end{align}

\end{document}